\documentclass[sort&compress,aps,a4paper,superscriptaddress,twocolumn]{revtex4}
\usepackage{amsmath}
\usepackage{graphicx}
\usepackage{color}
\usepackage{subfigure}
\usepackage{enumerate}
\usepackage{bm}
\usepackage{amssymb}
\usepackage{hyperref}

\newcommand{\hz}{\hat{\mathbf{z}}}

\newcommand{\bgeqa}{\begin{eqnarray}}
\newcommand{\edeqa}{\end{eqnarray}}
\newcommand{\vfluid}[1]{\textrm{\em {#1}}{}}	
\newcommand{\refeq}[1]{Eq.~(\ref{#1})}

\hypersetup{
		bookmarksopen=false,
    unicode=false,          
    pdftoolbar=true,        
    pdfmenubar=true,        
    pdffitwindow=true,     
    pdfstartview={FitH},    
    pdfpagemode=UseNone,		
    pdftitle={Dynamics of Ion Temperature Gradient Turbulence and Transport with a Static Magnetic Island},    
	pdfauthor={Olivier Izacard, Christopher Holland, Spencer D. James and Dylan P. Brennan},     
    breaklinks=true,
    pdfnewwindow=true,      
    colorlinks=true,       
    linkcolor=blue,          
    citecolor=blue,        
    filecolor=magenta,      
    urlcolor=blue           
}

\begin{document}
\title{Dynamics of Ion Temperature Gradient Turbulence and Transport with a Static Magnetic Island}
\author{O. Izacard}
\email{izacard@llnl.gov}
\affiliation{Lawrence Livermore National Laboratory, 7000 East Avenue, Livermore, California 94550, USA}
\author{C. Holland}
\affiliation{University of California - San Diego, La Jolla, California 92093-0417, USA}
\author{S. D. James}
\affiliation{University of Tulsa, Tulsa, Oklahoma 74104, USA}
\author{D. P. Brennan}
\affiliation{Princeton University, Princeton, New Jersey 08544, USA}

\begin{abstract}
Understanding the interaction mechanisms between large-scale magnetohydrodynamic instabilities and small-scale drift-wave microturbulence is essential for predicting and optimizing the performance of magnetic confinement based fusion energy experiments. We report progress on understanding these interactions using both analytic theory and numerical simulations performed with the BOUT++ [B. Dudson et al., Comput. Phys. Comm. {\bf 180}, 1467 (2009)] framework. This work focuses upon the dynamics of the ion temperature gradient instability in the presence of a background static magnetic island, using  a weakly electromagnetic two-dimensional five-field fluid model. It is found that the island width must exceed a threshold size (comparable to the turbulent correlation length in the no-island limit) to significantly impact the turbulence dynamics, with the primary impact being an increase in turbulent fluctuation and heat flux amplitudes.  The turbulent
radial ion energy flux is shown to localize near the X-point, but does so asymmetrically in the poloidal dimension.  An effective turbulent resistivity which acts upon the island outer layer is also calculated, and shown to always be significantly (10x - 100x) greater than the collisional resistivity used in the simulations.
\end{abstract}

\date{\today}
\maketitle

\section{Introduction}

One of the most crucial factors in determining the size of magnetic-confinement based fusion energy (MFE) reactors is the level of plasma confinement achieved \cite{Doyle:07}.  In MFE-relevant plasmas, the dominant physical processes which determine this confinement are a variety of instabilities driven by the inherent temperature, density, and current gradients of the confined plasma.  Two of the most important such instabilities in tokamaks are the ion temperature gradient (ITG) mode \cite{Horton_99_RMP} and tearing modes \cite{Furth:63}.  As its name suggests, the ITG instability is driven by the radial gradient (i.e. across nested magnetic flux surfaces) of the equilibrium ion temperature profile, which must exceed a critical value (the so-called critical gradient) for onset of the instability.  Once the critical gradient is exceeded, the ITG instability leads to a broad spectrum of ion gyroradius ($\rho_i$) scale fluctuations, whose collective nonlinear behavior drives rapid cross-field particle, energy, and momentum fluxes that greatly exceed collisional transport levels \cite{Helander:02}, and often determine the overall confinement level achieved.  Since these fluctuations generally have correlation lengths on the order of $1 - 10 \rho_i$, which is much smaller than the minor radius $a$ in MFE-relevant devices, the saturated fluctuations and their dynamics are often referred to as microturbulence.


Magnetic islands can appear through magnetohydrodynamic (MHD) instabilities~\cite{Freidberg:87, Biskamp:93}, such as the tearing mode (TM) driven by equilibrium gradients, or be externally imposed with resonant magnetic perturbations (RMP). These islands can reduce the confinement achieved, similar to the ITG instability, or even terminate confinement by leading to a disruption. 
It is theoretically predicted and experimentally observed that the presence of a magnetic island flattens the electron and ion temperature and density.  This flattening occurs both inside the island through rapid parallel equilibration, both through parallel sound waves~\cite{Scott:85} and rapid parallel heat conduction~\cite{Fitzpatrick:95}, and in time averaged profiles by enhancing the effective cross-field transport. 
 The classic picture is that the radial heat flux increases at the X-point of the magnetic island due to the fast parallel dynamics in the highly anisotropic 3D island structure \cite{Fitzpatrick:95}. Indeed, because of the direct connection of the magnetic flux surface at the separatrix of the island, all density and temperature structures on one side of the island can be rapidly transported to the other side of the island on shorter timescales than the timescale of the background cross-field transport. 

Given the importance of both classes of instabilities in determining the level of confinement achieved in magnetically confined plasmas, it is important to develop  models of their dynamics which can be used to accurately interpret existing observations and confidently design future experiments and reactors.  While great progress has been made in recent years in understanding the nonlinear dynamics and scalings of each class of instabilities in isolation (often through use of massively parallel computation), these instabilities frequently coexist in current high-performance tokamak discharges, and are expected to continue to do so in many future reactor scenarios.  We must therefore also understand how these instabilities couple linearly and nonlinearly if we are to be able to predict the overall confinement and performance of such discharges.  For example, it is now well-known that microturbulence (and ITG microturbulence in particular) generally saturates via the nonlinear generation of axisymmetric zonal flows \cite{Diamond_ZF_review_PPCF:05} which in turn shear apart the ITG eddies, forming a self-regulating system of turbulence and zonal flows.  How this process is altered by the presence of magnetic islands (which generally have toroidal mode numbers $n = 1-2$ and poloidal mode numbers $m = 2-5$) and their associated flow fields remains an open question.  Alternatively, the saturation of the neoclassical tearing mode depends sensitively on the level of profile flattening that occurs in and near the island \cite{Chang:95,Carrera:86}.  As this flattening depends sensitively on the ratio of parallel and perpendicular (microturbulence-dominated) transport, it is clear that the saturated island width will depend self-consistently on the level of microturbulence present (which also depends on the amount of profile flattening that occurs).

These two instabilities (ITG and TM) are well studied in the literature, with many different research teams finding a wide array of complex nonlinear couplings between microturbulence and tearing modes. For instance, McDevitt and Diamond \cite{McDevitt:06} demonstrated that drift-wave microturbulence could nonlinearly excite an island via the Reynolds stress in a fashion analogous to the formation of zonal flows, while Sen \textit{et al} \cite{Sen:09} used a similar approach to demonstrate that electron temperature gradient (ETG) modes could nonlinearly damp the island via an effective turbulent resistivity.  
More recently, a wide range of researchers have utilized both fluid 
\cite{
Ishizawa:07a,
Ishizawa:07b,
Militello:08, 
Waelbroeck:09a, 
Waelbroeck:09b, 
Muraglia:09, 
Muraglia:11, 
Ishizawa:10a,
Ishizawa:10b,
Ishizawa:13, 
Agullo:14,
Hu:14,
Hill:15}
 and gyrokinetic 
 \cite{
 Wilson:09,
 Peeters:09,
 Poli:09,
 Poli:10,
 Hornsby:11,
 Siccinio:11,
 Waltz:12,
 Liu:14,
 Zarzoso:15,
 Hornsby:15a,
 Hornsby:15b}
 approaches to investigate couplings of various microturbulence instabilities to either statically imposed islands or dynamically evolving tearing modes.  Although there is no simple and clear consensus yet, it is clear that the presence of the island can lead to a poloidal localization of linear microturbulence eigenmodes, and that the microturbulence can both damp and destabilize islands depending on the type of microturbulence instability, and the width and rotation speed of the island.  The impacts of the turbulence on the island evolution are generally represented by inclusion of turbulent polarization current terms in generalized Rutherford equations \cite{Rutherford:73, Waelbroeck:09b}.
 
Building upon these results, we present in this paper the findings of a study examining the responses of weakly electromagnetic ITG turbulence to statically imposed islands of varying width, performed using a simple fluid model in slab geometry.  We find that above a critical island width, the magnitude of both the microturbulence fluctuations and associated radial ion heat flux $Q_i$ increase approximately linearly with island width, both near and far from the island resonant surface, for values of the the ion temperature gradient both close to, and well above the linear critical gradient.  The critical island width is found to be approximately equal to the radial correlation length of the turbulence in the absence of an island. Perhaps not surprisingly, for islands smaller than the critical size, the microturbulence is found to be effectively insensitive to the presence of the island.  
We also observe the turbulent $Q_i$ to localize near the island X-point as expected, but somewhat surprisingly, this localization is not seen to be symmetric in the poloidal direction.
To begin assessing how the turbulence may be expected to back-react upon the island, we present calculations of an effective turbulent resistivity which operates on the ``outer solution" of the island structure.  We show that for the parameters considered, while this resistivity is always much larger than the collisional value, it maximizes for small island size and decreases rapidly with increasing island width.

The paper is organized as follows. Section~\ref{ToC.Theo} details the fluid model used to study the effects of a static magnetic island on the ITG microturbulence, and includes details of the numerical implementation in BOUT++ (\ref{ToC.numerical_implementation}) and verification of the implementation using analytic growth rate calculations (\ref{ToC.linear_verification}) and a global energy balance analysis suitable for nonlinear simulations (\ref{ToC.energy_balance_verification}).  Sec.\ref{ToC.IslandOnTurb} discusses the results of numerical simulations quantifying the impact of statically imposed islands of varying width on ITG turbulence, and Sec.\ref{ToC.EtaTurb} discusses initial work exploring the expected feedback of the turbulence on the island.  Conclusions and future research directions are discussed in Sec.\ref{ToC.Future}


\section{Details of the five-field fluid model}
\label{ToC.Theo}
This section describes the 2D electromagnetic five-field model used in this study, and its implementation in the BOUT++ framework \cite{Dudson_2009_CPC_180}.
Two verification methods are presented to test the implementation.  The first is a comparison of calculated linear growth rates against analytic derivations, detailed in Sec. \ref{ToC.linear_verification}.  The second is a global energy balance analysis approach, detailed in Sec. \ref{ToC.energy_balance_verification}.

\subsection{Model overview}
In order to investigate the self-consistent couplings of ITG turbulence and tearing modes in a simple, tractable form, we employ a simple five-field two-fluid model~\cite{Ishizawa:10a,Hazeltine_1985_PoF_28,Ishizawa:12} in a Cartesian slab geometry~\cite{Ishizawa:10a,Fitzpatrick:05a,Fitzpatrick:05b,Fitzpatrick:05c,Ishizawa:12} similar to that used in previous studies, considered together only in Ishizawa~\textit{et al.}~\cite{Ishizawa:10a,Ishizawa:12}. 
The simulation radial and poloidal domains span the range $-a \le x \le a$ and $-b/2 \le y < b/2$ respectively, with Dirichlet boundary conditions on all fluctuating fields at $x = \pm a$ and periodic boundary conditions at $y=\pm b/2$.
Using the standard definitions of
$\vec{E} = -\vec{\nabla} \phi$ and $\vec{B} = B_0(x) \hz + \vec{\nabla} \psi \times \hz$, the model describes the evolution of fluctuations in the
electron density $\tilde{n} =\delta n/\bar{n}$,
vorticity $\tilde{\Omega} = \nabla_{\perp}^2 \tilde{\phi}$,
ion parallel velocity $\tilde{v}_{i,z} = \delta v_{i,z}/c_s$,
ion temperature $\tilde{T}_i = \delta T_i/\bar{T}$,
and magnetic flux $\tilde{\psi} = (1/\beta)(c_s/c) e \delta \psi/\bar{T}$, 
where $\tilde{\phi} = e\delta \phi/\bar{T}$ 
is the normalized electrostatic potential fluctuation, and the normalized axial current is given by ${j}_z = - \nabla_{\perp}^2 {\psi}$.  
Here, $\bar{n}$ and $\bar{T}$ are constant normalizing factors taken to be equal to the values of the equilibrium density and ion temperature profiles at $x=0$; $c_s = \sqrt{\bar{T}/M_i}$ and $\rho_s = c_s/\Omega_{ci}$, with $\Omega_{ci} = eB_0(0)/M_i c$.
In this model, we can chose a form of the equilibrium magnetic flux that allows us to specify both an 
equilibrium poloidal magnetic field $B_{y,0}(x) = - \partial_x \psi_0$ and a static magnetic island $B_{x,island}(y) = \partial_y \psi_{island}$. These relations come from $\bm{B}= \bm{\nabla} \psi(x,y) \times \hz$ and the current is given by $j_{z0} = \hz \cdot \bm{\nabla} \times \bm{B} = - \nabla^2 \psi$.
To further simplify the problem, in the rest of this paper we restrict our attention to the case for which the equilibrium
density $n_0$ is constant in radius and the equilibrium electrostatic potential $\phi_0$ and parallel ion flow $v_{i,z0}$ are equal to zero.  The equilibrium ion
temperature profile is taken to have a simple linear dependence $T_{i0}(x) = \bar{T}(1 - x/L_{Ti})$, where $L_{Ti}$ is constant across the radial domain.  
The equilibrium magnetic flux is taken to have the form $\psi_{eq}(x,y) = \psi_0(x) + \psi_{island}(y) = -x^2/(2 L_S) + \psi_i \cos(2 \pi y/b)$, with $\psi_i = k_0 W_{island}^2 / (4a)$ and the associated current is $j_{z0}(x,y) = - \nabla_{\perp}^2 \psi_{eq}$ where $\nabla_{\perp}^2 = \partial_{xx}^2 + \partial_{yy}^2$.  The constituent equations of the model are the electron density, the perpendicular momentum associated with the quasi-neutrality, the parallel momentum, the ion temperature conservation, and Ohm's law as follows:
\bgeqa
\label{eq.dndt}
\displaystyle
d_t \tilde{n} &=& \vfluid{v}_B^{\star} \partial_y \left( \tilde{\phi}-\tilde{n} \right) + \nabla_{\parallel} \left( \tilde{j}_z - \tilde{v}_{i,z} \right) 
+ \tilde{\nabla}_{\parallel} j_{z0} \nonumber\\
\displaystyle
&-& \nu_{s}\left< \tilde{n} \right>_{y} + D_{\perp} \nabla_{\perp}^2 \tilde{n}\\
\label{eq.dvordt}
\displaystyle
d_t \tilde{\Omega} &=& \nabla_{\parallel} \tilde{j}_z + \tilde{\nabla}_{\parallel} j_{z0} \\
&-& \vfluid{v}_{B}^{\star} \partial_y \{(1+\tau) \tilde{n} + \tau \tilde{T}_i\}   + \tau \vfluid{v}_{T_i}^{\star} \partial_y \tilde{\Omega} \nonumber\\
\displaystyle
&+&\tau \bm{\nabla}_{\perp} \cdot \left[ \bm{\nabla}_{\perp} \tilde{\phi} , \tilde{n} + \tilde{T}_i \right]  + D_{\perp} \nabla_{\perp}^2 \tilde{\Omega} \\
\label{eq.dvpardt}
\displaystyle
d_t \tilde{v}_{i,z} &=& - \nabla_{\parallel} \left( (1+\tau) \tilde{n} + \tau \tilde{T}_i \right) \nonumber\\
\displaystyle
&-& \nu_{s}\left< \tilde{v}_{i,z} \right>_{y} + D_{\perp} \nabla_{\perp}^2 \tilde{v}_{i,z} \\
\label{eq.dTidt}
\displaystyle
d_t \tilde{T}_i &=& - \vfluid{v}_{T_i}^{\star} \partial_y \tilde{\phi} - (\Gamma-1) \nabla_{\parallel} \tilde{v}_{i,z}  \nonumber\\
\displaystyle
&+&\lambda \vfluid{v}_B^{\star} (\Gamma -1) \partial_y \left( \tilde{\phi}/\tau+\tilde{n}+\tilde{T}_i\right)   \nonumber\\
\displaystyle
&-&  \nu_{s} \left< \tilde{T}_i \right>_{y}  + D_{\perp} \nabla_{\perp}^2 \tilde{T}_i \\
\label{eq.dpsidt}
\displaystyle
\beta \partial_t \tilde{\psi} &=& -\nabla_{\parallel} \left(\tilde{\phi}-\tilde{n}\right) - \eta \tilde{j}_z
\edeqa
where $v_B^{\star}=a/L_B$ and $v_{Ti}^{\star}=a/L_{Ti}$ are the driven imposed gradients of the magnetic field and the ion temperature. 
In these equations $x$ and $y$ are normalized to $\rho_s$ and $t$ to $a/c_s$. Using the standard Poisson bracket notation $\left\{f,g\right\} = \partial f/\partial x \, \partial g/\partial y - \partial f/\partial y \, \partial g/\partial x$, 
we write the total time derivative as $df/dt = \partial f/\partial t + \left\{\tilde{\phi},f \right\}$ to include the $\bm{E} \times \bm{B}$ convection term, and the total parallel derivative
$\nabla_{\parallel}f = -\beta \left\{\psi_0 + \psi_{island} + \tilde{\psi}, f \right\}$ includes derivatives along the total magnetic field (the equilibrium field, any imposed island, and
the magnetic field fluctuations)
. We also note that the use of the notation $\tilde{\nabla}_{\parallel}f = -\beta \left\{ \tilde{\psi}, f \right\}$ is associated only with the magnetic fluctuations part.
The coefficient $\Gamma=5/3$ is the ratio of specific heats and $\tau$ is the ratio between the ion and electron reference temperatures. 
%
The coefficient $\lambda$, equal to $0$ or $1$, is used to investigate the impact of different closure models in the $\tilde{T}_i$ equation. We have included $\lambda$ in order to investigate the effects of the absence in some models found in the literature of the divergence of the ion $\bm{E}\times\bm{B}$ and diamagnetic velocity terms. This term, describing finite Larmor-radius (FLR) effects, is retained (or not) in the fluid closure model used for
the ion temperature equation. Moreover, this model contains other FLR effects (i.e., $\tau \bm{\nabla}_{\perp} \cdot \left[ \bm{\nabla}_{\perp} \tilde{\phi} , \tilde{n} + \tilde{T}_i \right]$)~\cite{Hsu_1986_PoF_29,Brizard:92,Scott_2000_PoP_7} which are not always considered in the literature but included here in order to take into account the ion temperature effects which come from the convection of the ion pressure by the $\bm{E} \times \bm{B}$ drift. This term appears in the vorticity equation due to the divergence of the ion polarization current. When the inhomogeneity of the magnetic field is kept, the models including this FLR term such as in Refs.~\cite{
Hsu_1986_PoF_29,
Brizard:92,
Hazeltine_1985_PoF_28,
Hazeltine:87,
Miyato_2004_PoP_11,
Ishizawa_2007_PoP_14,
Ishizawa_2013_PoP_20} 
are not Hamiltonian~\cite{Izacard_2011_PoP_18} in the ideal limit ($D=\eta=\nu_s=0$). 
However, the ideal version of our model is very close to a Hamiltonian structure because the terms required to recover the Hamiltonian property involve, for example, higher order spatial derivatives of the magnetic field which are canceled by the assumption of a linear background magnetic gradient, or higher order spatial derivatives of the vorticity which are assume negligible. Moreover, the energy conservation of our simulations is systematically verified as well as the error between the numerical and theoretical computation of the energy balance. 
We found that the inclusion of the FLR term (with $\lambda=1$) changes the growth rate less than $10\%$ in the linear regime, but this FLR term in the ion temperature equation and the one in the vorticity equation are kept in the following because they contribute to the numerical stability of our simulations for high $k_y$ modes.

This set of equations becomes an electrostatic two-dimensional four-field fluid model with gradient driven turbulence when $\beta=0$. Moreover, when $\beta$ is relatively small (i.e., $\sim 10^{-5}$) the differences between the five-field and the four-field electrostatic model are not significant. The five-field model is preferred for consistency with future work on dynamic islands. The additional dissipative viscosity $\nu_s$ is included to prevent quasilinear relaxation and maintain input background profiles such as the background ion temperature gradient. Without this forcing, the ion temperature gradient cannot be maintained in the presence of a magnetic island and the ITG which drives the turbulence becomes stable. This choice is needed to maintain on average the same level of turbulence due to the ITG. We show below that this assumption is still compatible with the flattening effect of the island on the temperature and density profiles. 
The numerical dissipation terms proportional to $D_{\perp} \nabla_{\perp}^2$ utilize small values of $D_{\perp}$ to damp grid-scale fluctuations without significantly impacting the large-scale dynamics of interest.  Note that relative to previous tearing mode studies which have utilized anomalous transport coefficients, these simulations self-consistently determine those coefficients via the turbulent fluxes e.g. $\bm{Q}_i = \left< \tilde{T}_i {\tilde{\bm{v}}_{E \times B}}\right>$.

\subsection{Numerical implementation}
\label{ToC.numerical_implementation}
The BOUT++~\cite{Dudson_2009_CPC_180} framework is used for the numerical computation of the five-field fluid model given by Eqs.~(\ref{eq.dndt})-(\ref{eq.dpsidt}). Even though BOUT++ can deal with a helical axisymmetric magnetic field, as a first step our work focuses on a slab geometry choosing the box ratio of the simulations in order to fix a specific regime of the instabilities. The use of BOUT++ is an asset since the dynamical equations are written in a small script with object oriented functions. Also, all multi-dimensional scans and post-treatment analyses have been simplified by the use of the OMFIT~\cite{Meneghini_2013_PFR_8} framework and has motivated the creation and development of the BOUT++ interface module in the OMFIT framework as explained in Ref.~\cite{Meneghini_2015_NF_IAEA}. 
As a summary, OMFIT is a software (with Graphical User Interface) for integrated studies which can quickly connect many codes and experimental data in a workflow and which can perform data analysis. 
For instance, OMFIT was used to handle remote execution, data transfer, and analysis of the simulations presented here, which were performed mainly on the TSCC computer at the San Diego Supercomputer Center.

\subsection{Linear dispersion relation}
\label{ToC.linear_verification}
In order to verify the accuracy of our numerical simulations against the target fluid model, 
the first step is to compare the computed linear growth rates against analytic calculations in the limit of 
no background static island and no static magnetic shear (i.e., $\psi_{eq}=0$).
The linearization of the fluid equations by $\tilde{f}(x,y,z,t) = \tilde{f} \exp(ik_xx+ik_yy-i\omega t)$ yields 
\bgeqa
\displaystyle
- \omega \tilde{n} &=& \vfluid{v}_B^{\star} k_y \left( \tilde{\phi}-\tilde{n} \right),\\
\displaystyle
\omega k_{\perp}^2 \tilde{\phi} &=& - \vfluid{v}_{B}^{\star} k_y ((1+\tau) \tilde{n} + \tau \tilde{T}_i) \nonumber\\
\displaystyle
&&
- \tau \vfluid{v}_{T_i}^{\star} k_y k_{\perp}^2 \tilde{\phi}, \\
\displaystyle
- \omega \tilde{v}_{i,z} &=& 0, \\
\displaystyle
- \omega \tilde{T}_i &=& - \vfluid{v}_{T_i}^{\star} k_y \tilde{\phi} \nonumber\\
\displaystyle
&&
+ \vfluid{v}_B^{\star} k_y \lambda (\Gamma -1) \left( \tilde{\phi}/\tau+\tilde{n}+\tilde{T}_i\right), \\
\displaystyle
-\omega \beta \tilde{\psi} &=& i \eta k_{\perp}^2 \tilde{\psi}.
\edeqa
After some analytic computation 
the dispersion relation is reduced to
\bgeqa
\label{eq.DispRel}
\displaystyle
D(\omega) = A_5 \omega^5 + A_4 \omega^4 + A_3 \omega^3 + A_2 \omega^2 + A_1 \omega + A_0,
\edeqa
with
\bgeqa
\displaystyle
A_5 &=& \beta k_{\perp}^2, \\
\displaystyle
A_4 &=& k_{\perp}^2 ( i \eta k_{\perp}^2 + \beta C_{BTi} ), \\
\displaystyle
A_3 &=& - \beta \Omega_B ( (1+\tau+\Gamma_2) \Omega_B - \tau \Omega_{Ti} ) 
+ k_{\perp}^4 i \eta C_{BTi} \nonumber\\
\displaystyle
&&
- \beta k_{\perp}^2 \Omega_B ( \Gamma_2 \Omega_B - \tau (\Gamma_2-1) \Omega_{Ti} ), \\
\displaystyle
A_2 &=& - i \eta k_{\perp}^2 ( (1 + \tau + \Gamma_2) \Omega_B^2 - \tau \Omega_{Ti} \Omega_{B} - k_{\perp}^2 C_{BTi} ) \nonumber\\
\displaystyle
&&
- \beta \tau \Omega_{Ti} \Omega_{B}^2 ( \Gamma_2 k_{\perp}^2 + 1 ), \\
\displaystyle
A_1 &=& k_{\perp}^2 i \eta ( \Gamma_2 k_{\perp}^2 + 1 ) \tau \Omega_{Ti} \Omega_{B}^2, \\
\displaystyle
A_0 &=& 0,
\edeqa
where $\Omega_B=v_B^{\star} k_y$, $\Omega_{Ti}=v_{Ti}^{\star} k_y$, $\Gamma_2=\lambda (\Gamma-1)$ and $C_{BTi} = (\Gamma_2-1) \Omega_B + \tau \Omega_{Ti}$.

This linear dispersion relation is obviously driven by the imposed gradients of the ion temperature $v_{Ti}^{\star}$ and magnetic field $v_B^{\star}$. If the latter are dropped, the dispersion relation becomes $D(\omega)=k_{\perp}^2 \omega^4 (\beta \omega + i\eta k_{\perp}^2)$ and the roots are found to be stable (imaginary part of $\omega$, i.e. the growth rate, is $0$ and the propagation frequency is equal to $0$ or $- k_{\perp}^2 \eta/\beta$). A scan of the ion temperature and magnetic field gradients has shown that the linear growth rate is unstable beyond a threshold dependent on both gradients. We typically perform our simulations in the unstable regime with $v_B^{\star}=0.1$ and $v_{Ti}^{\star}=0.3$, highly relevant to experimental values.

\begin{figure}[!ht]
\centering
\subfigure[\ Growth rate $\gamma$]{
\includegraphics[height=30mm]{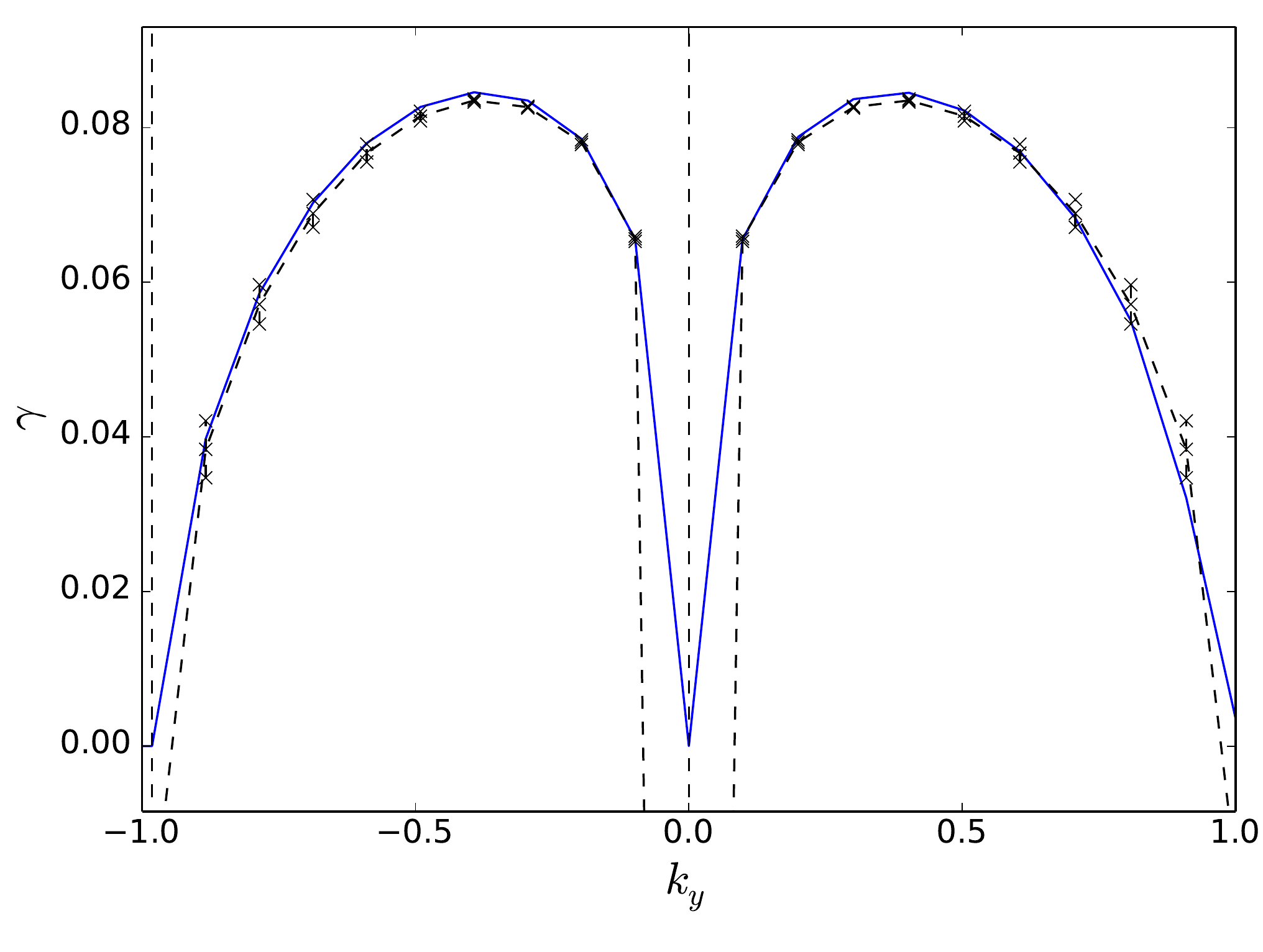}
}
\subfigure[\ Propagation $\omega$]{
\includegraphics[height=30mm]{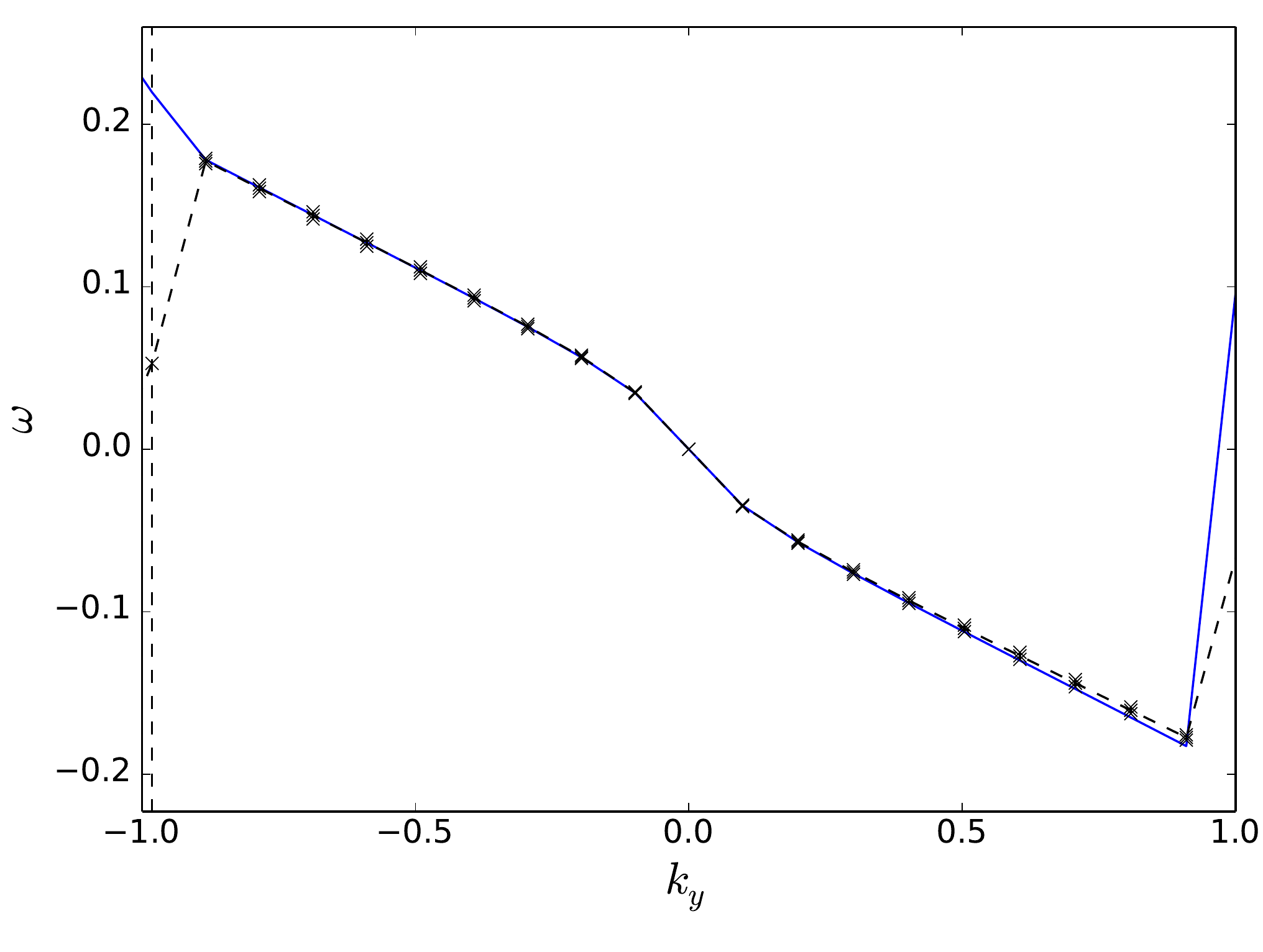}
}
\caption{Verification of the linear dispersion relation computation between the theory (blue curves) given by the Eq.~(\ref{eq.DispRel}) and the BOUT++ simulation (dashed-black curves) of the Eqs.~(\ref{eq.dndt})-(\ref{eq.dpsidt}). The error bars (x-markers) of the simulation are plotted. The error bar becomes artificially big when the growth rate is stable or very small in the unstable regime.}
\label{fig.DispRel}
\end{figure}

A comparison of the computed linear growth rates against the analytic linear dispersion relation (Eq.~(\ref{eq.DispRel})) is shown in Fig.~\ref{fig.DispRel},
demonstrating excellent agreement in both growth rate $\gamma$ and real frequency $\omega$ for this base case; similar agreement is obtained at other parameters.  
Error estimates on the simulation results are calculated via the standard deviation of the instantaneous frequency and growth rate of the time-averaging window used (generally on the order of $100$'s of $a/c_s$).  For wave numbers with sufficiently small or zero growth rate, a well-converged value is often not found, leading to the large error bars shown at the highest values of $k_y$ plotted and at $k_y=0$.

\subsection{Energy evolution and energy balance}
\label{ToC.energy_balance_verification}
\begin{figure}[!hbps]
\centering
\includegraphics[width=85mm]{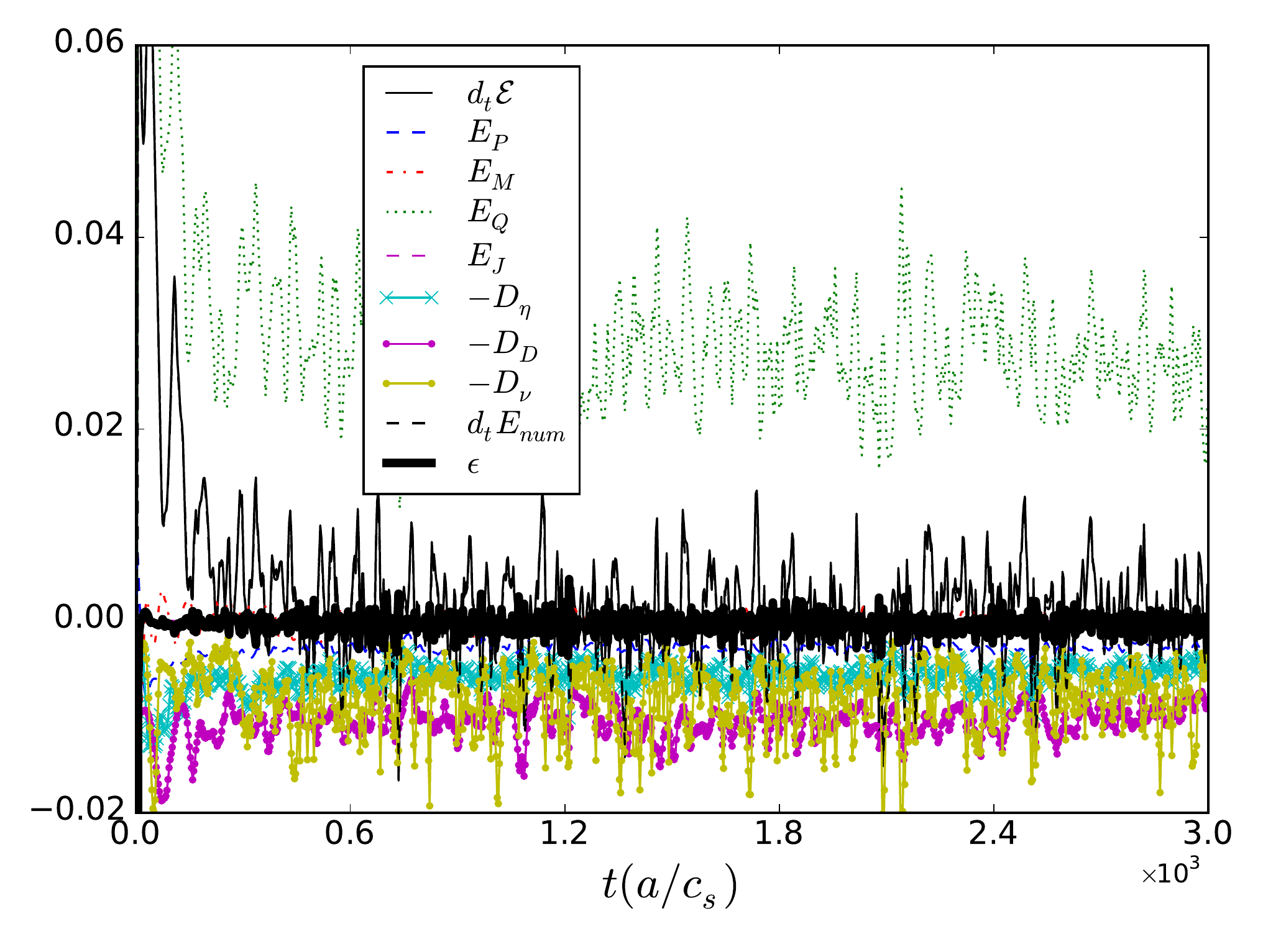}
\caption{Time evolution of the energy balance in arbitrary units with a static background magnetic island of width $W_{island}=15 \rho_s$. Each term of Eq.(\ref{eq.dEdt}) is represented in color as well as their sum $d_t\mathcal{E}$ (plain black curve) and the numerical value $d_t E_{num}$ (dashed-black curve). The relative difference $\epsilon$ defined by \refeq{eq.dEdt_error} between the two later is negligible (bold black curve).}
\label{fig.dEdt}
\end{figure}

\begin{figure}[!hbps]
\centering
\includegraphics[width=85mm]{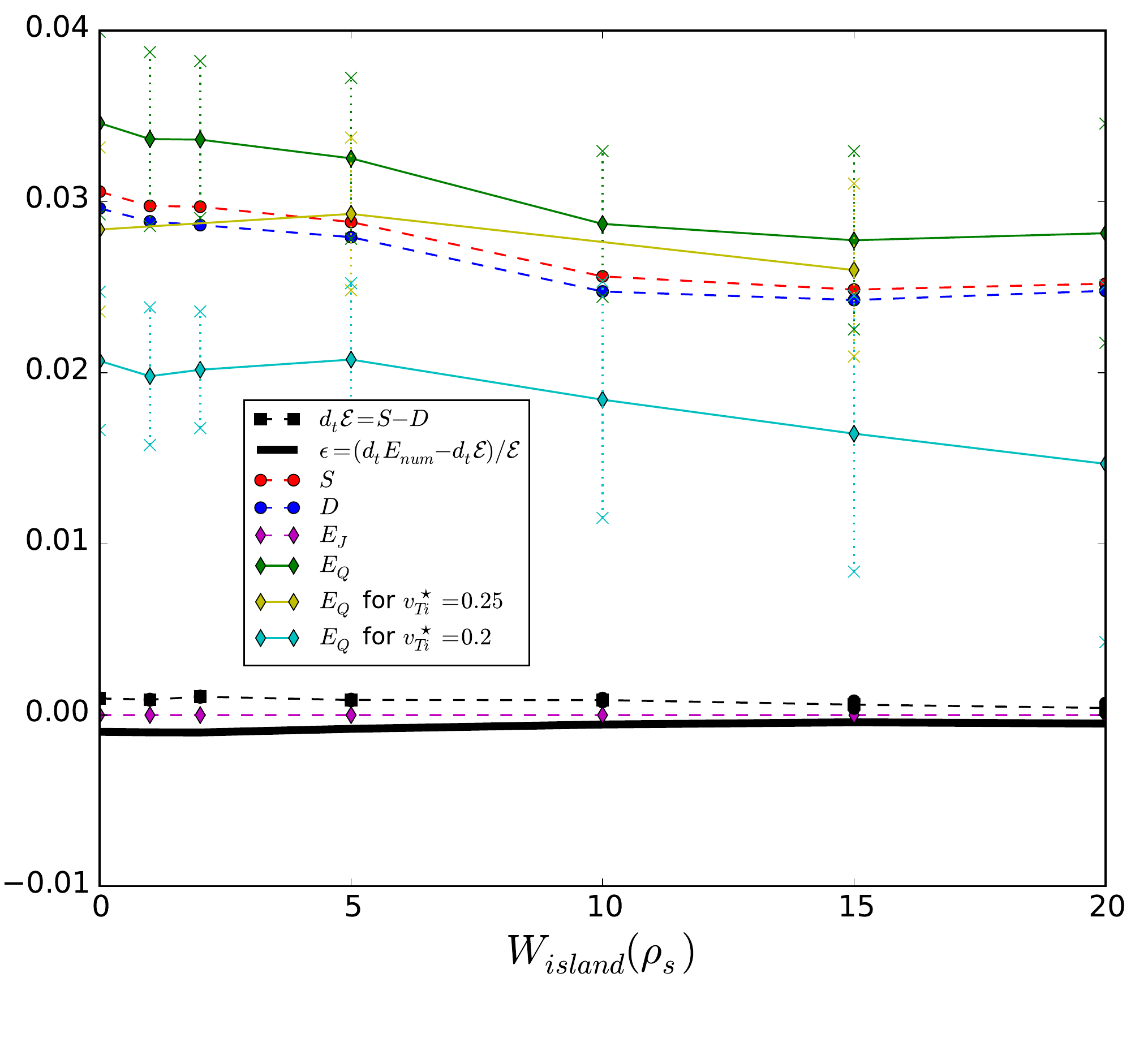}
\caption{Scaling in arbitrary units of time-averaged terms in the energy balance analysis as a function of $\vfluid{v}_{T_i}^{\star} = a/L_{Ti}$ over the time range $t \in \left[ 500, 3000 \right] a/c_s$ in the presence of islands of widths $W_{island}=\{0,1,2,5,10,15\}$, and magnetic island size at fixed $\vfluid{v}_{T_i}^{\star} = \{0.2,0.25,0.3\}$. The sum of the positive source (respectively negative dissipative) terms of the energy balance given by~\refeq{eq.dEdt} are noted $S$ (respectively $D$). The error bars correspond to the standard deviation.}
\label{fig.dEdt_avg}
\end{figure}

The next step of the verification process is to quantify the global energy balance of the simulations, which provides an error measure when the analytic calculation of the linear dispersion relation cannot be used, due to inclusion of a non-uniform background magnetic field and/or nonlinear effects.  The total energy of the model given by Eqs.~(\ref{eq.dndt})-(\ref{eq.dpsidt}) is
\bgeqa
\label{eq.E}
\displaystyle
\mathcal{E} &=& \frac{1}{2} \int d^2x \bigg[ (1+\tau) \left( \tilde{n}^2 + \vert \nabla_{\perp}\tilde{\phi} \vert^2 
+ \beta \vert \nabla_{\perp}\tilde{\psi} \vert^2 \right) \bigg. \nonumber\\
\displaystyle
\bigg. &&\qquad
+ \tilde{v}_{i,z}^2 + \frac{\tau}{\Gamma-1} \tilde{T}_i^2 \bigg],
\edeqa
where the first term corresponds to the internal potential energy of the ions and electrons, the second term to the perpendicular kinetic energy, the third term to the magnetic fluctuation energy, the fourth term to the parallel kinetic energy and the final term to the ion thermal energy.  
After some analytic computations 
the time evolution of the total energy is
\bgeqa
\label{eq.dEdt}
d_t \mathcal{E} = E_P + E_M + E_Q + E_J - D_{\eta} - D_D - D_{\nu},
\edeqa
by specifying all components of the energy balance
where the individual terms are given by
\bgeqa
\displaystyle
E_P &=& \lambda \tau \vfluid{v}_B^{\star} P, \\
\displaystyle
E_M &=& \tau (1+\tau) \vfluid{v}_B^{\star} M, \\
\displaystyle
E_Q &=& \left[ \left\{ \tau (1+\tau) - \lambda \right\} \vfluid{v}_{B}^{\star} + \frac{\tau}{\Gamma-1} \vfluid{v}_{T_i}^{\star} \right] Q, \\
\displaystyle
E_J &=& (1+\tau) \left< \left(\tilde{n} - \tilde{\phi}\right) \tilde{\nabla}_{\parallel} j_{z0}(x,y) \right>_{x,y}\\
\displaystyle
D_{\eta} &=& (1+\tau) \eta \left< \tilde{j}_z^2 \right>_{x,y}, \\
\displaystyle
D_D &=& D_{\perp} \bigg<  (1+\tau) \left( \vert \nabla_{\perp} \tilde{n} \vert^2 + \vert \nabla_{\perp}^2 \tilde{\phi} \vert^2 \right) \bigg. \nonumber\\
\displaystyle
\bigg. & & + \vert \nabla_{\perp} \tilde{v}_{i,z} \vert^2 + \frac{\tau}{\Gamma-1} \vert \nabla_{\perp} \tilde{T}_i \vert^2 \bigg>_{x,y} \\
\displaystyle
D_{\nu} &=& \nu_s \bigg( (1+\tau) \left< \tilde{n} \right>_{x,y}^2 \bigg. \nonumber\\
\displaystyle
\bigg. &&
+ \left< \tilde{v}_{i,z} \right>_{x,y}^2 + \frac{\tau}{\Gamma-1} \left< \tilde{T}_i \right>_{x,y}^2 \bigg),
\edeqa
with the positive driven terms related to: a pressure gradient flux $P = \left< \tilde{T}_i \partial_y \tilde{n} \right>_{x,y}$, the radial particle flux $M = \left< \tilde{n} \tilde{v}_x \right>_{x,y}$, the radial ion heat flux $Q = \left< \tilde{T}_i \tilde{v}_x \right>_{x,y}$ due to the $\bm{E} \times \bm{B}$ drift~\footnote{The component of the radial ion heat flux due to the $\nabla B$-drift is not taken into account here.} where $\tilde{v}_x = -\partial_y \tilde{\phi}$, and the negative dissipative terms related to: the resistivity $\eta$, the numerical diffusion $D_{\perp}$ and the axisymmetric sink terms proportional to $\nu_s$ which are used for the conservation of the background radial profiles. The coupling of the fluctuations to background axial and island current gradients $E_J$ is added in order to verify its contribution in the energy balance. 
However, our choice of magnetic shear $B_{y0}(x)$ does not contribute to $E_J$ since its second derivative vanishes, and so for the cases considered here, $E_J$ contains only the contribution of the imposed island current gradient.  In Fig.~\ref{fig.dEdt} and~\ref{fig.dEdt_avg} below it is shown to be negligible for all island widths relative to the dominant source and sink terms.

Each of these terms can be calculated in the simulations and the energy conservation properties of a given simulation can thereby be quantified in detail. The terms $E_f$ are in general positive drives ($E_J$ can be positive or negative), and the terms $D_f$ are dissipative contributions. 
As mentioned previously, we note that for the equations for $E_P$ and $E_Q$, the presence of the FLR term $\lambda=1$ stabilizes the turbulence. In fact, the magnetic gradient part of the energy balance related to the heat flux $E_Q$ is divided by $2$ (when $\tau=1$) and Fig.~\ref{fig.dEdt_avg} shows a negative contribution of $E_P$ due to the presence of the FLR term.
By calculating each of these terms from the simulation output, a numerical error measure $\epsilon$ can be readily defined as
\bgeqa
\label{eq.dEdt_error}
\displaystyle
\epsilon = \frac{ d_t E_{num} - d_t \mathcal{E} }{\mathcal{E}}.
\edeqa

For the nonlinear simulations we are using $125$ radial grid points and $128$ poloidal grid points with poloidal periodicity and radial Dirichlet boundary conditions. The size of the box is $(2a,b)=(65\rho_s,65\rho_s)$ so the discretization is $(dx,dy) \sim (0.52 \rho_s,0.50 \rho_s)$. There are no significant effects of increasing the number of grid points by a factor 2 in each direction. With these parameters, the microturbulence is well described and the saturation state is reached after a few hundreds of time steps ($a/c_s$) as shown below. 

Fig.~\ref{fig.dEdt} shows the time evolution of the accuracy of the energy balance between the analytic prediction given by Eq.~(\ref{eq.dEdt}) and the numerical computation. All components of the analytic prediction of the energy balance are shown in different colors. We can clearly see that the dominant component of the energy balance is the one related to the radial ion heat flux, as expected in our ITG driven model. The numerical computation of the energy balance given by the time derivative of the numerical total energy is almost exactly identical to the analytic prediction. The relative error between the analytic and numerical computation of the energy balance $\epsilon$ given by \refeq{eq.dEdt_error} is negligible with respect to the dominant term $E_Q$. We do not show here but, as expected, the part of the energy related the ion temperature fluctuations is dominant with respect to the other terms in Eq.~(\ref{eq.E}). This explains the dominance of $E_Q$ with respect to the other components $E_P$, $E_M$ and $E_J$. 
The values of the dissipative terms ($D=0.02$, $\nu_s=0.5$, $\eta=10^{-4}$, and $\beta=10^{-5}$) act on the saturation of the numerical simulations. Indeed in this nonlinear simulation, with the chosen parameters, the saturation is obtained after approximately 400 $a/c_s$ time units since the energy balance fluctuates around zero for a static island of width $W_{island}=15\rho_s$.  

In Fig.~\ref{fig.dEdt_avg}, time-averages (over the steady-phases) of normalized energies and energy balances $d_t \mathcal{E}$ with varying island width and the ion temperature gradient $\vfluid{v}_{T_i}^{\star}$ are shown, demonstrating the maintenance of good conservation for the simulation results discussed in the following sections. All curves correspond to an ion temperature gradient of $\vfluid{v}_{T_i}^{\star}=0.3$ except the yellow and cyan curves which are the dominant term of the energy balance $E_Q$ for $\vfluid{v}_{T_i}^{\star}=0.2$ and $0.25$. The numerical error is shown by the difference $d_t\mathcal{E}=S-D$ between the source $S=E_P+E_M+E_Q+E_J$ and dissipative terms $D=D_{\eta}+D_D+D_{\nu}$ which is near $0$. In particular, the value of the standard deviation of the theoretical prediction of the energy balance (${\rm std}(d_t\mathcal{E})$) averaged over the island width is $\approx 5\times 10^{-3}$ which is much smaller than the standard deviation of the dominant term (${\rm std}(E_Q)$) averaged over the island width $\approx 10^{-4}$. Moreover, the theoretical prediction of the energy balance $d_t\mathcal{E}$ is not $0$ but is around $10^{-3}$, which is much smaller than the dominant term $E_Q \approx 0.03$. Saturated energies increase with island width as discussed in Sec.~\ref{ToC.IslandOnTurb}, however the dependence of the normalized energy and energy balance on island width is weak, and the numerical error is low.

\section{Quantifying the impact of static island on ITG turbulence}
\label{ToC.IslandOnTurb}

In our simulations we force the system to maintain on average the same level of turbulence due to the ITG by keeping an average background ion temperature gradient. The electrons are assumed to be correlated with the ions and are not the focus of this work. The time evolution of the magnetic island is known to be at least one order of magnitude slower than the time evolution of the microturbulence driven by the ITG. An electrostatic model ($\beta=0$) allowing the exact force balance of the fluctuations in the Ohm's law between the parallel classical resistive current and the parallel gradient of the drifts (i.e., the electric and diamagnetic drifts) is not used in this paper. Instead, weakly electromagnetic (small $\beta=10^{-5}$) cases are considered, close to the electrostatic cases but, preparing for the full electromagnetic self-consistency simulations and allowing comparisons with other published electrostatic results. The advantage of full electromagnetic simulations is that the effects of the turbulence on the slow dynamical evolution of the magnetic island will be self-consistently included without using the artificially large numerical perpendicular transport coefficient.\\



\begin{figure}[!ht]
\centering
\includegraphics[width=85mm]{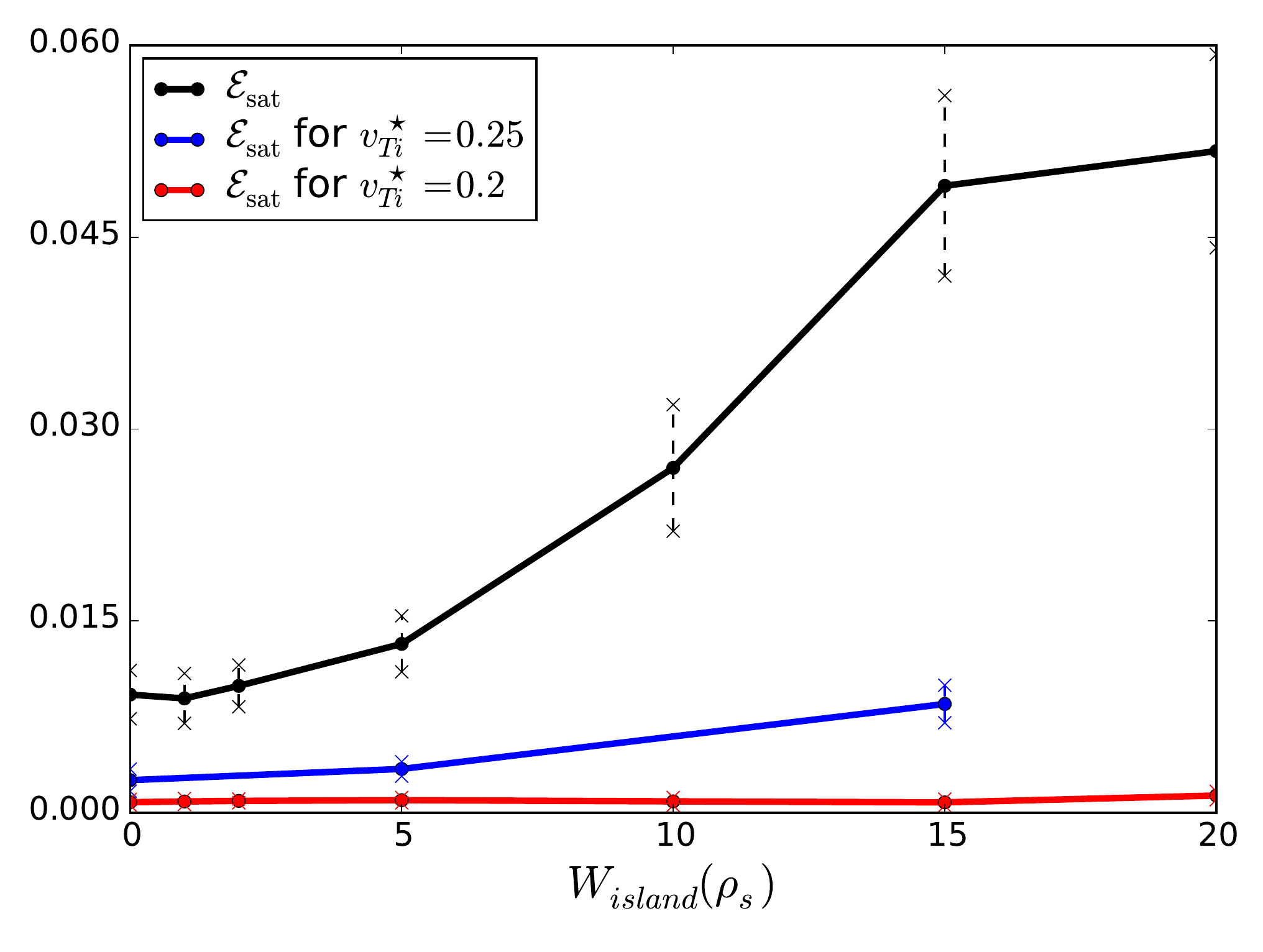}
\caption{Energy saturation in arbitrary units for different island sizes averaged over the time range $t \in \left[ 500; 3000 \right] a/ c_s$ for an ion temperature gradient $v_{Ti}^{\star}=0.3$ (black curve), $0.25$ (blue curve) and $0.2$ (red curve). The error bars correspond to the standard deviation.}
\label{fig.Esat.vs.Wisland}
\end{figure}

\begin{figure}[!htbp]
\centering
\subfigure[\ $W_{island}=0$]{
\label{fig.Ti_2D.a}
\includegraphics[width=40mm]{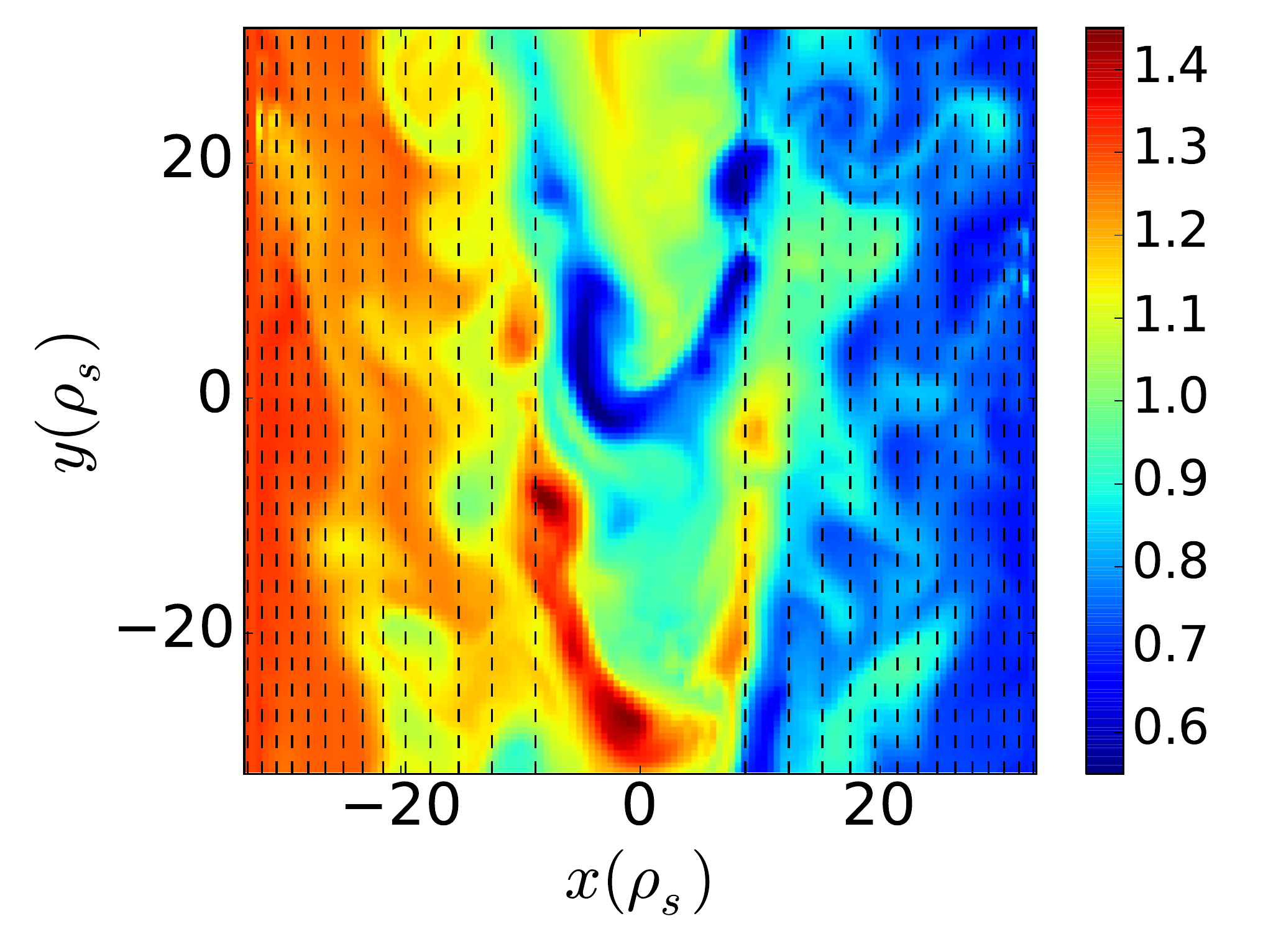}
}
\subfigure[\ $W_{island}=15\rho_s$]{
\label{fig.Ti_2D.b}
\includegraphics[width=40mm]{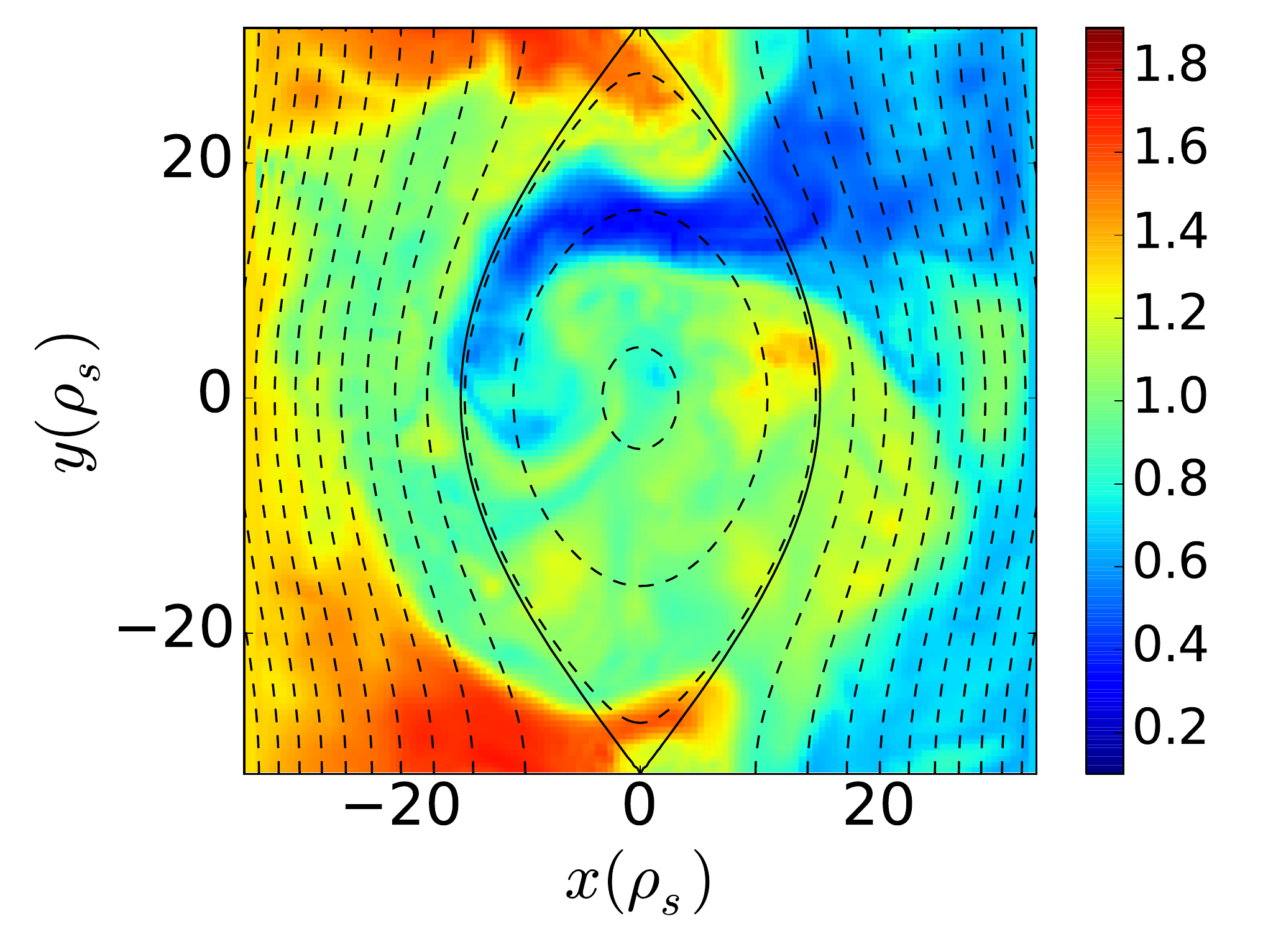}
}
\subfigure[\ $W_{island}=0$]{
\label{fig.Ti_2D.c}
\includegraphics[width=40mm]{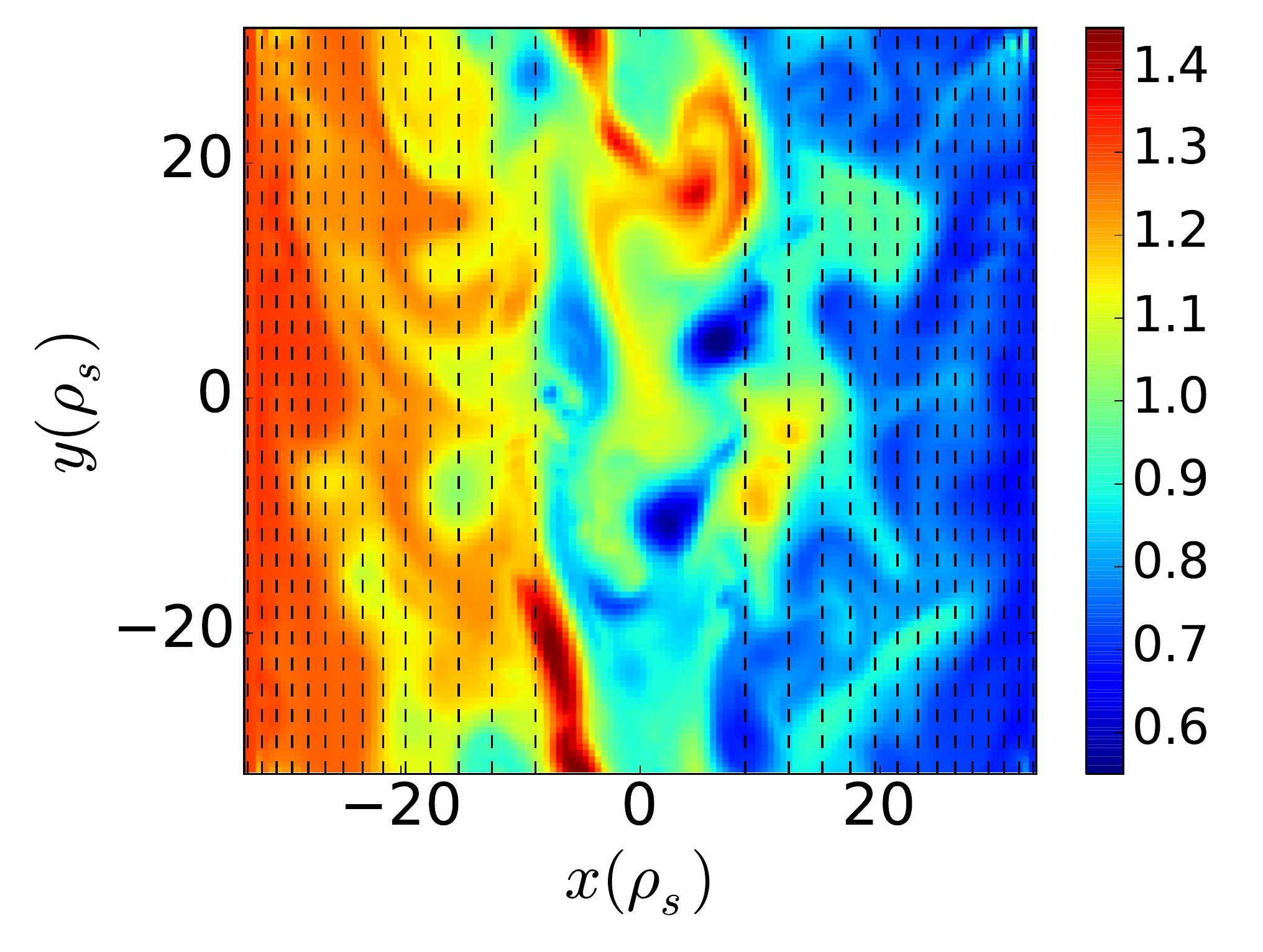}
}
\subfigure[\ $W_{island}=15\rho_s$]{
\label{fig.Ti_2D.d}
\includegraphics[width=40mm]{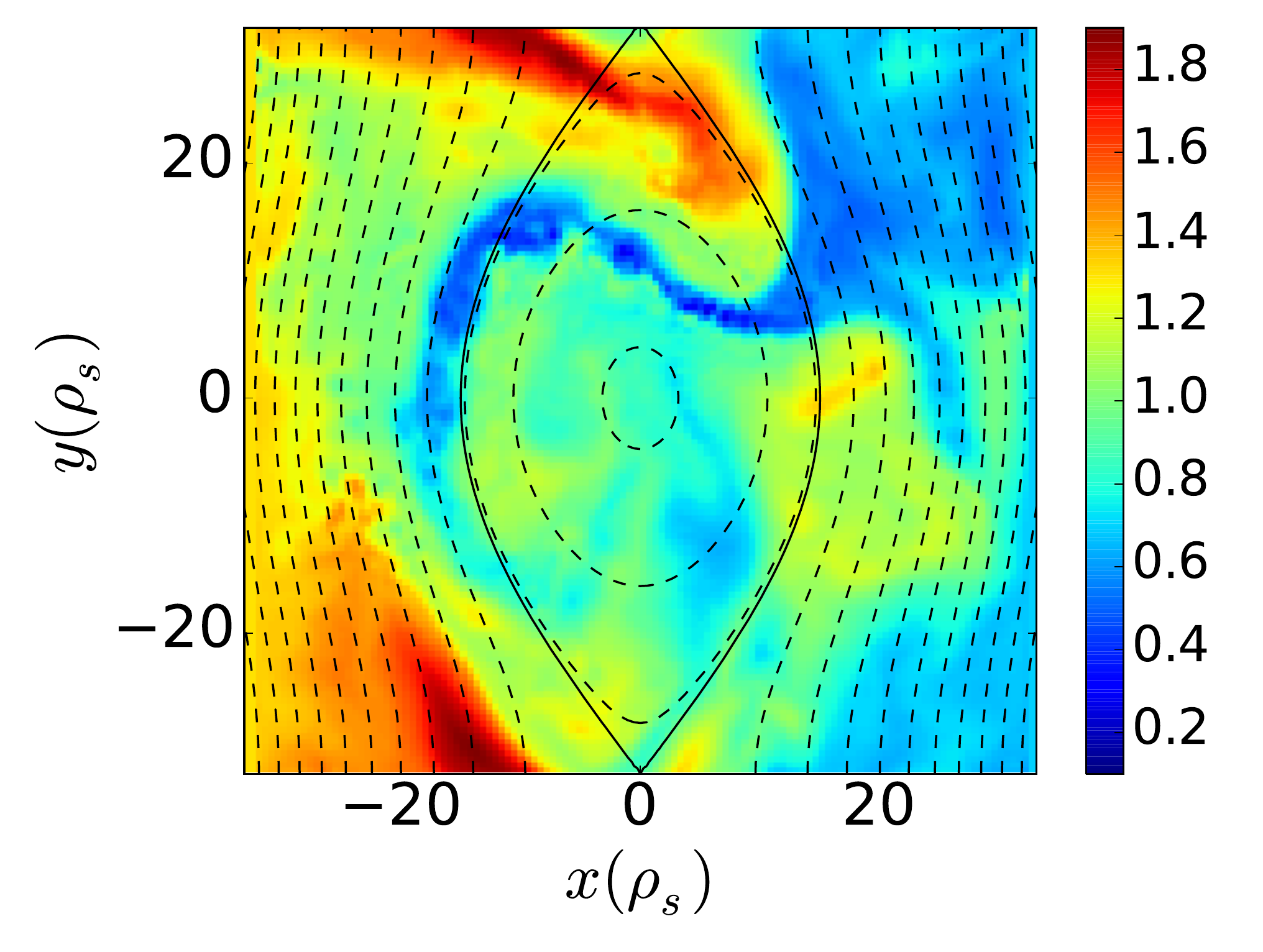}
}
\subfigure[\ $W_{island}=0$]{
\label{fig.Ti_2D.e}
\includegraphics[width=40mm]{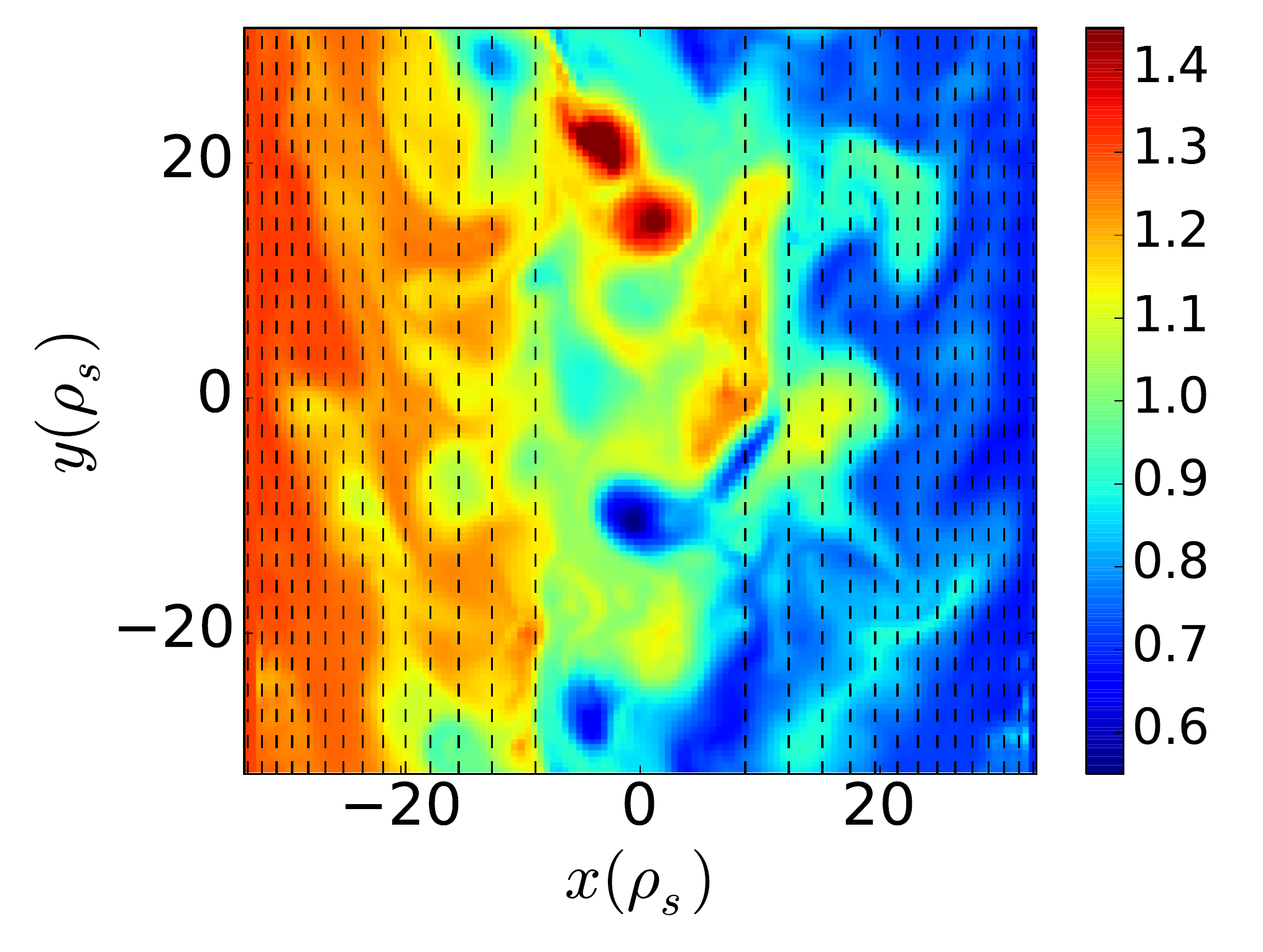}
}
\subfigure[\ $W_{island}=15\rho_s$]{
\label{fig.Ti_2D.f}
\includegraphics[width=40mm]{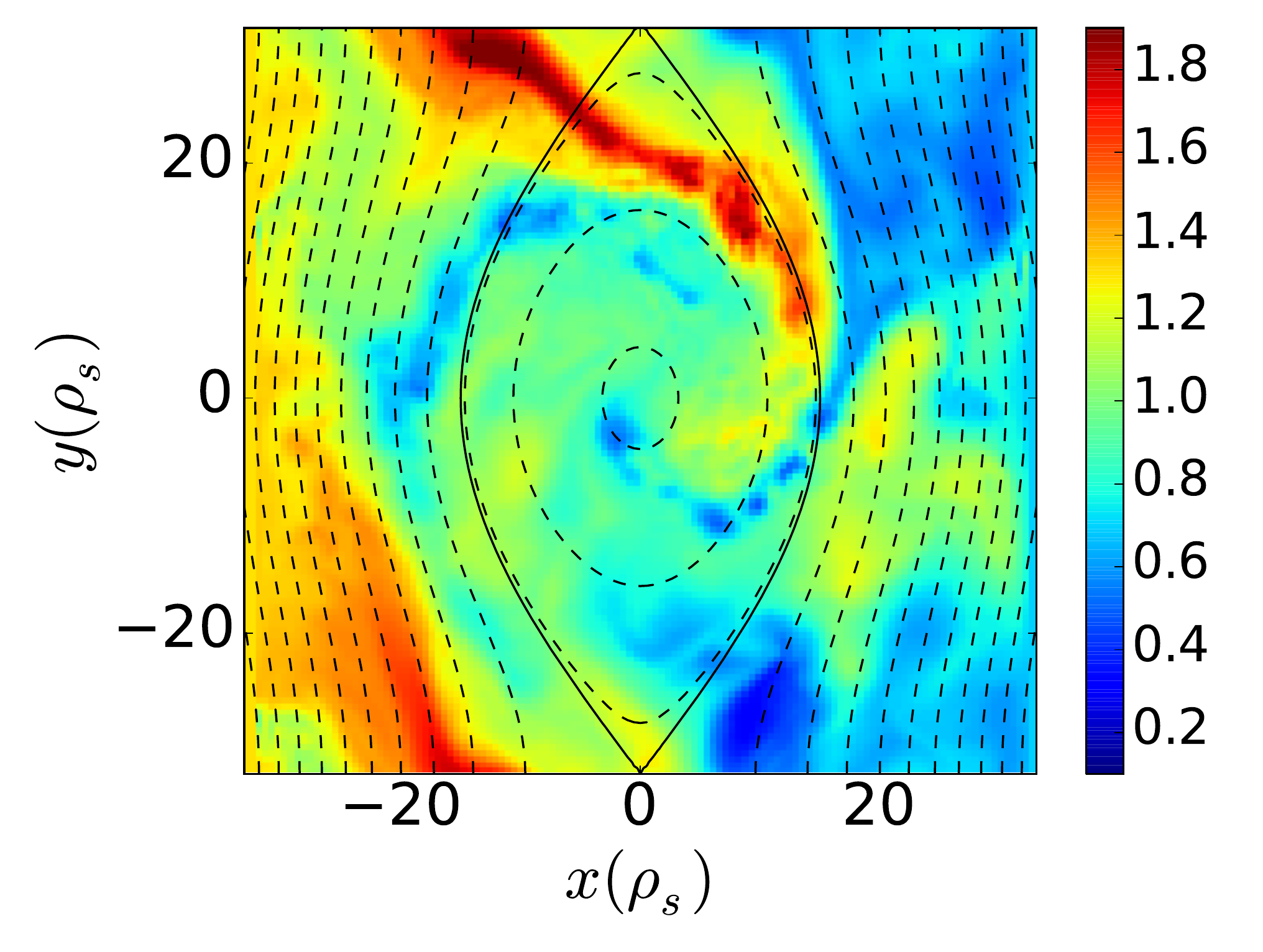}
}
\subfigure[\ $W_{island}=0$]{
\label{fig.Ti_2D.g}
\includegraphics[width=40mm]{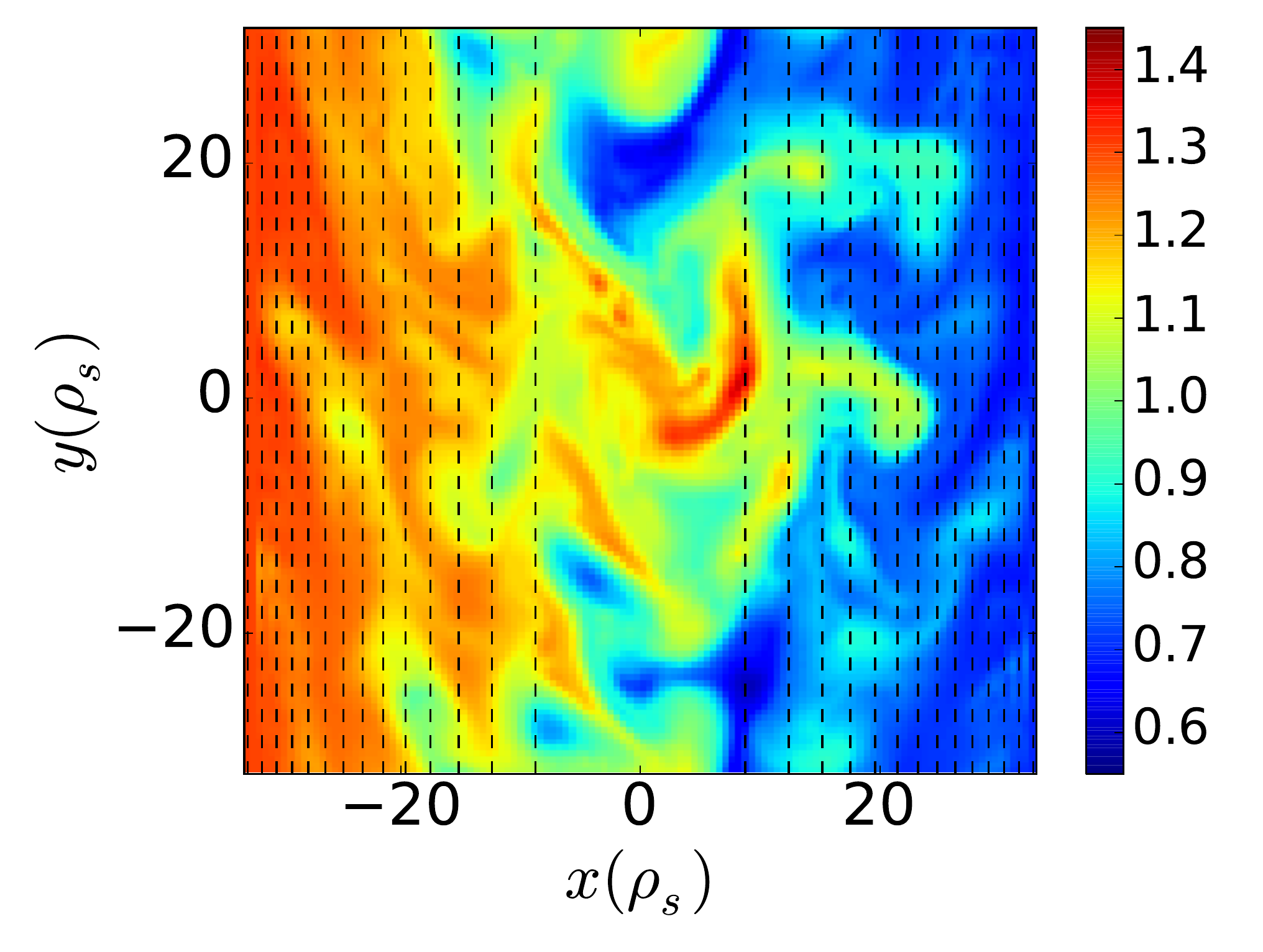}
}
\subfigure[\ $W_{island}=15\rho_s$]{
\label{fig.Ti_2D.h}
\includegraphics[width=40mm]{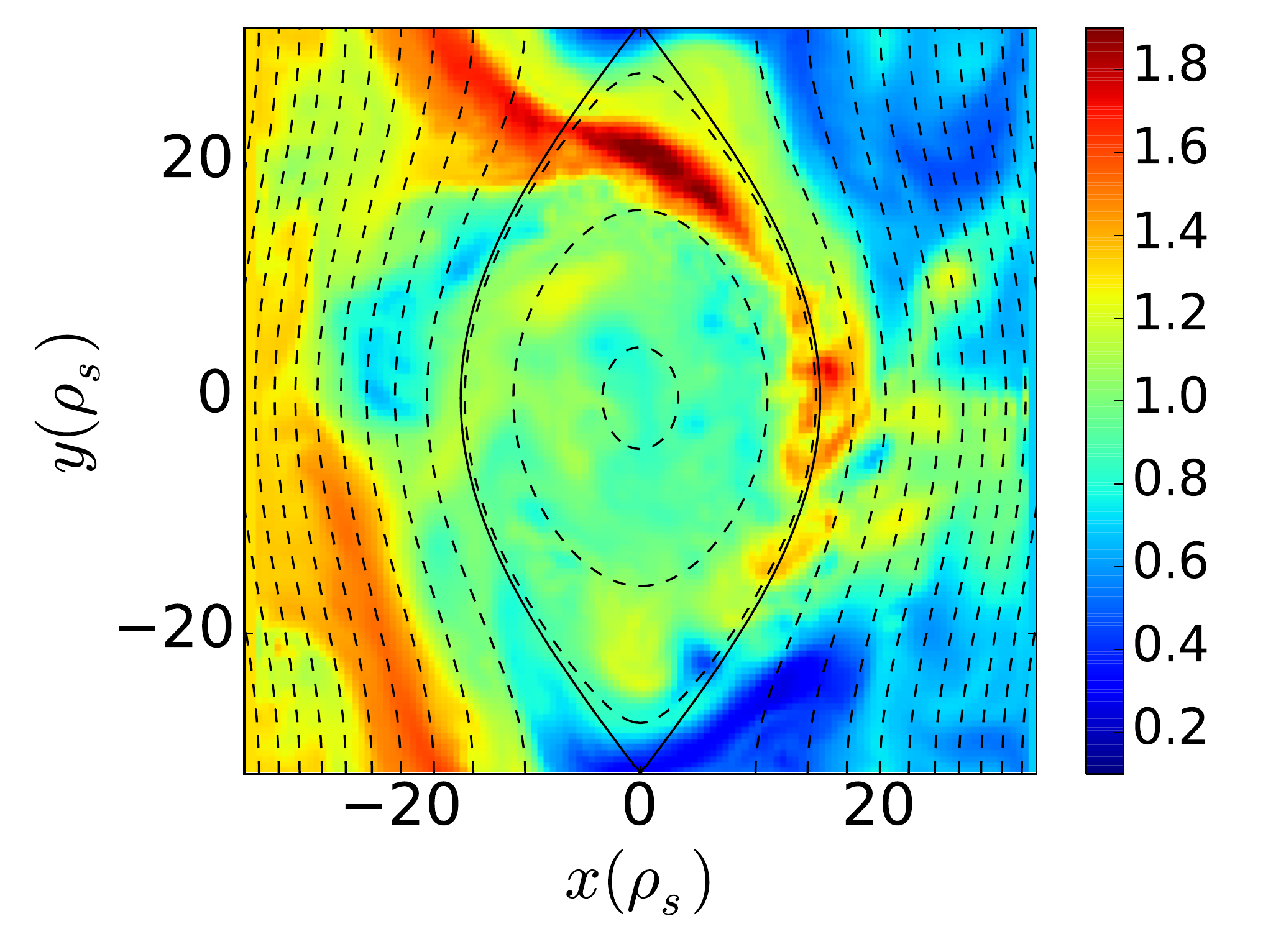}
}
\subfigure[\ $W_{island}=0$]{
\label{fig.Ti_2D.i}
\includegraphics[width=40mm]{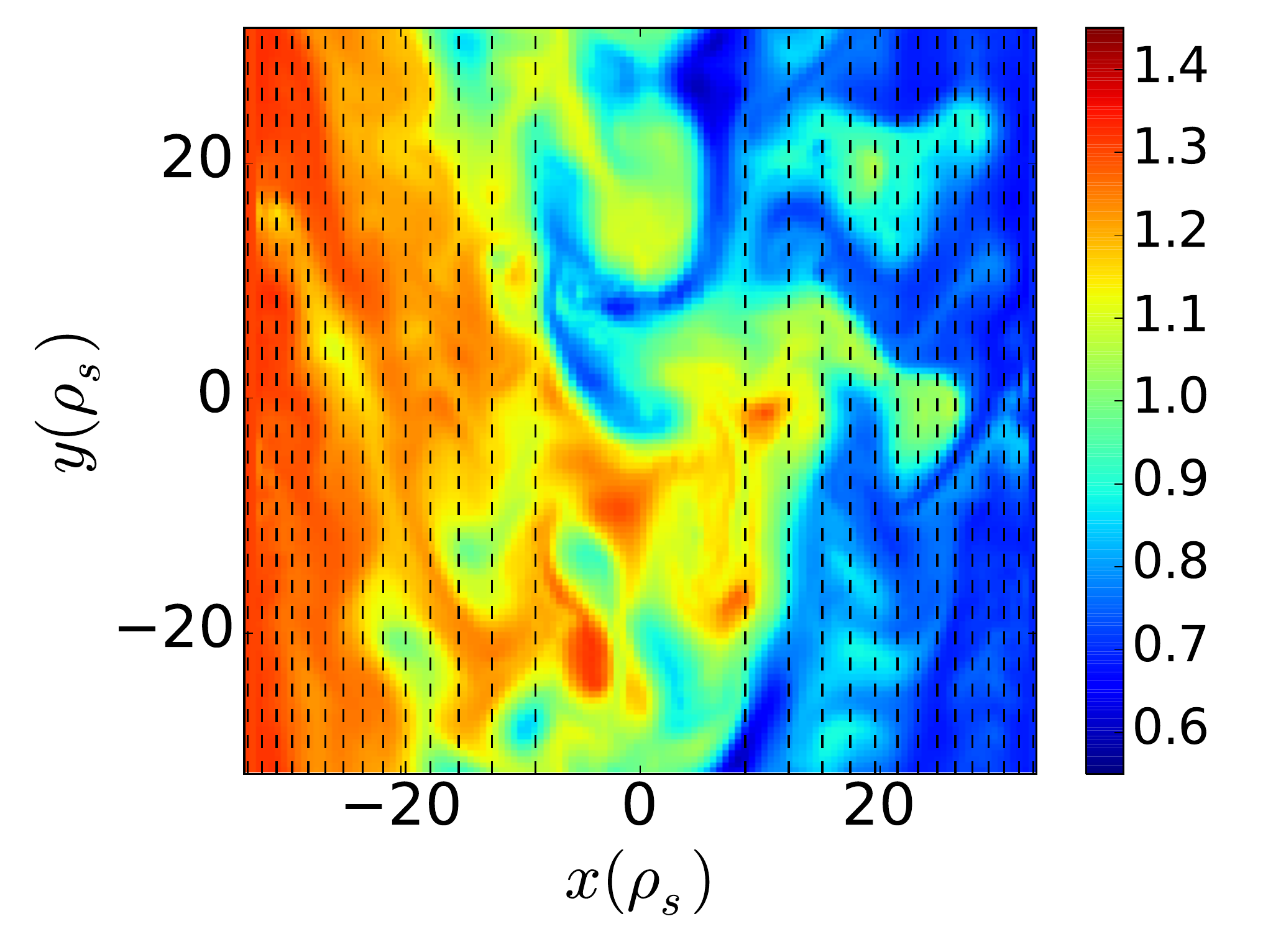}
}
\subfigure[\ $W_{island}=15\rho_s$]{
\label{fig.Ti_2D.j}
\includegraphics[width=40mm]{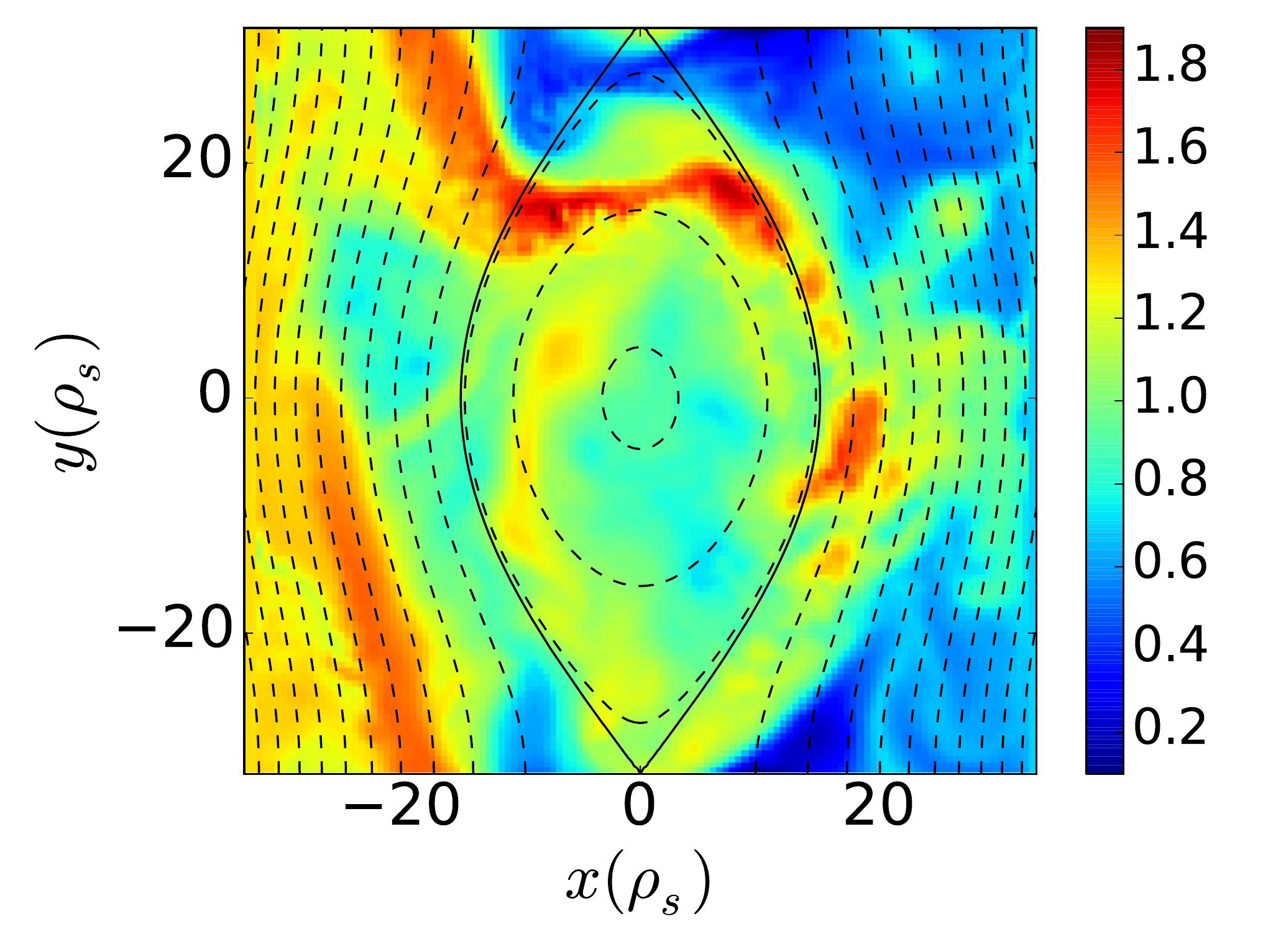}
}
\caption{Snapshots of the total ion temperature $T_i=T_{i0}(x)+\tilde{T}_i(x,y)$ at $t=\{ 1506, 1516, 1526, 1536, 1546 \} a / c_s$ respectively for (a)-(b), (c)-(d), (e)-(f), (g)-(h) and (i)-(j), obtained from two simulations with $W_{island}= 0 \rho_s$ for (a), (c), (e), (g), (i) and with $W_{island}= 15 \rho_s$ for (b), (d), (f), (h), (j). The same colorbar limits (i.e., $\left[0.55;1.45\right]$ and $\left[0.1;1.9\right]$) are used for all time steps of these two simulations (i.e.,  $W_{island}= 0 \rho_s$ and $W_{island}= 15 \rho_s$).}
\label{fig.Ti_2D}
\end{figure}

\begin{figure}[!htbp]
\centering
\subfigure[\ $W_{island}=0$]{
\includegraphics[width=40mm]{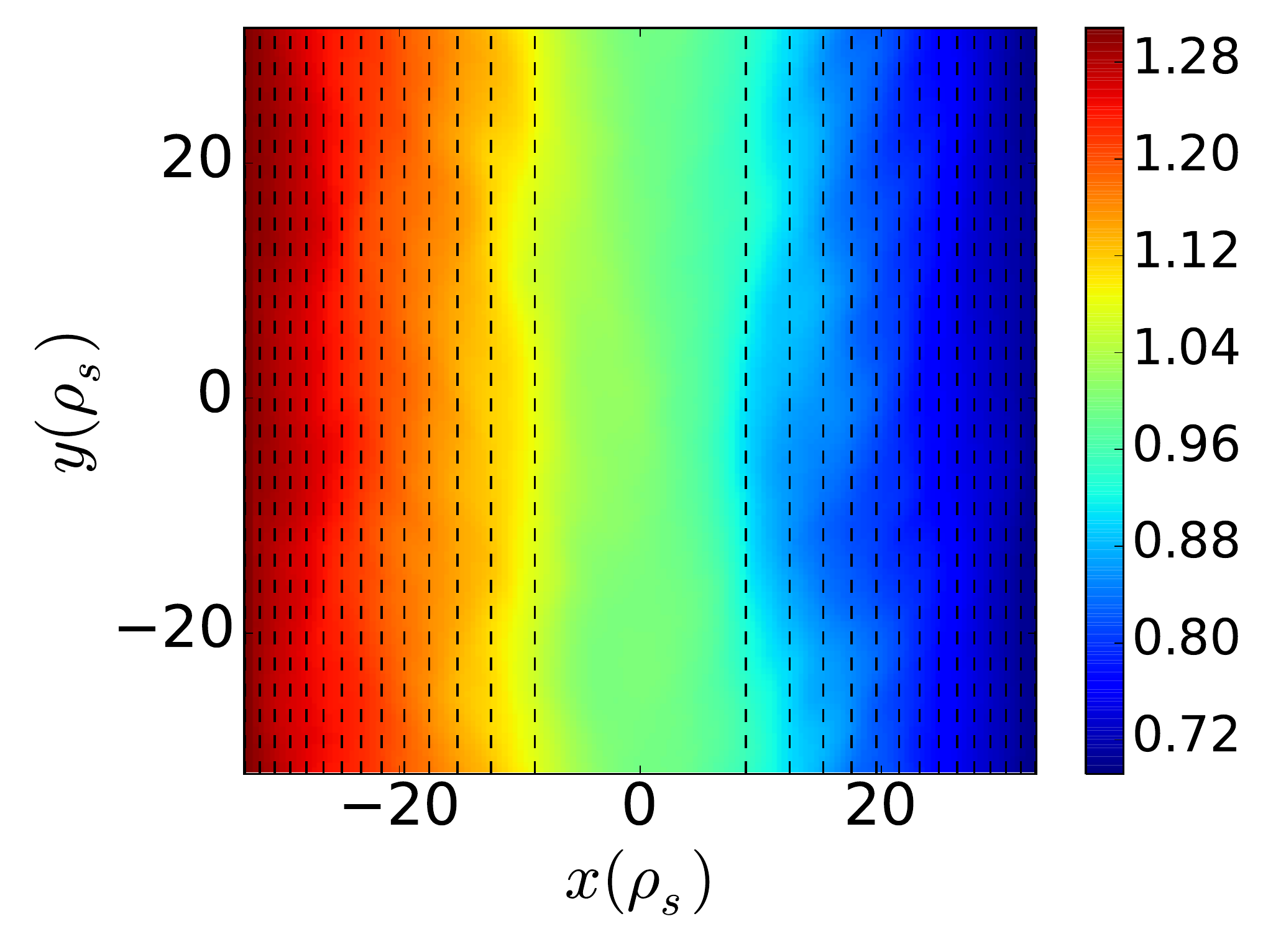}
}
\subfigure[\ $W_{island}=15\rho_s$]{
\includegraphics[width=40mm]{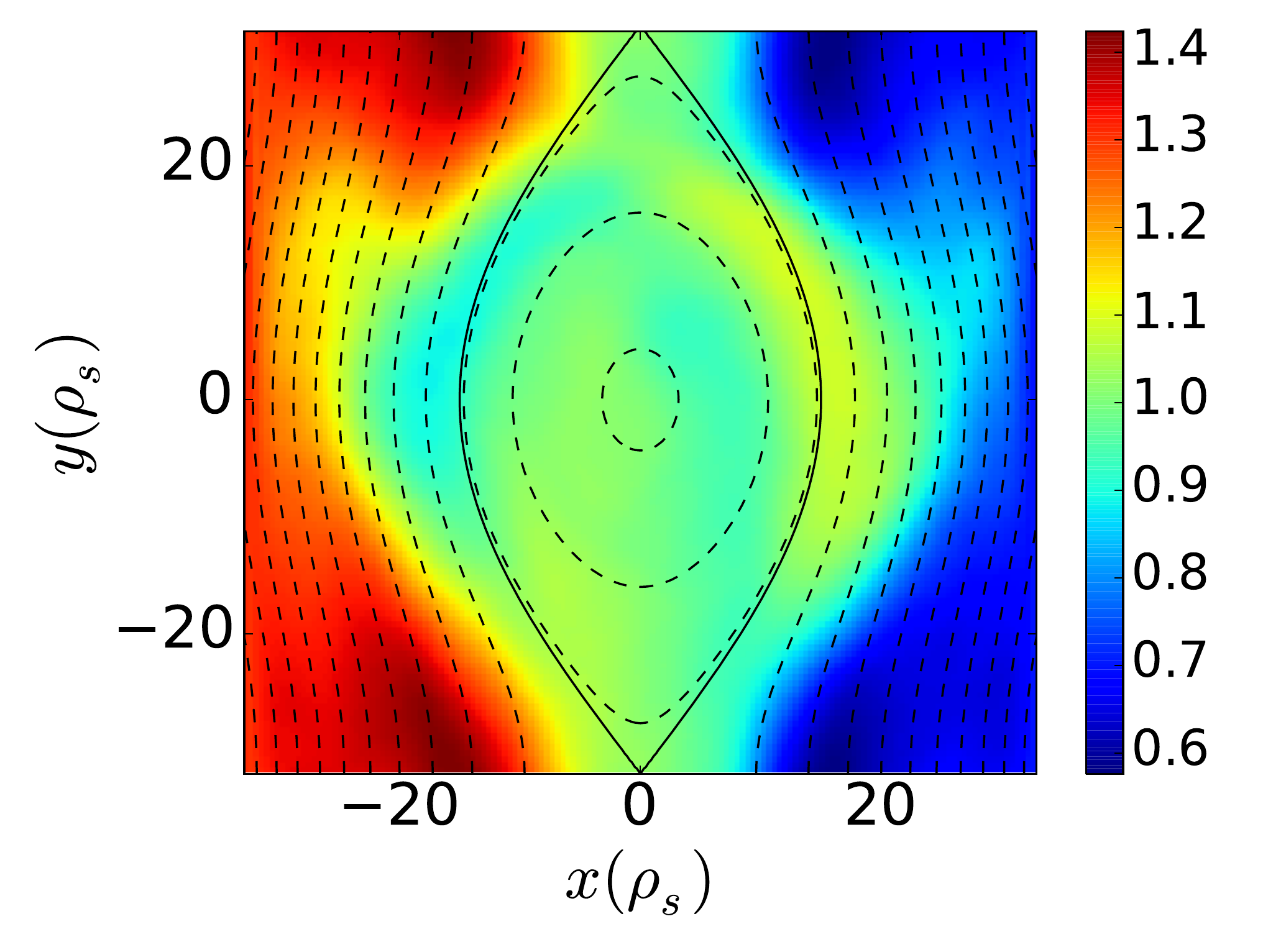}
}
\caption{Time average on $t \in [500;3000] a / c_s$ of the total ion temperature $T_i=T_{i0}+\tilde{T}_i$ for (a) $W_{island}=0\rho_s$ and (b) $W_{island}=15\rho_s$.}
\label{fig.Ti_2D_avg}
\end{figure}

\begin{figure}[!ht]
\centering
\includegraphics[width=85mm]{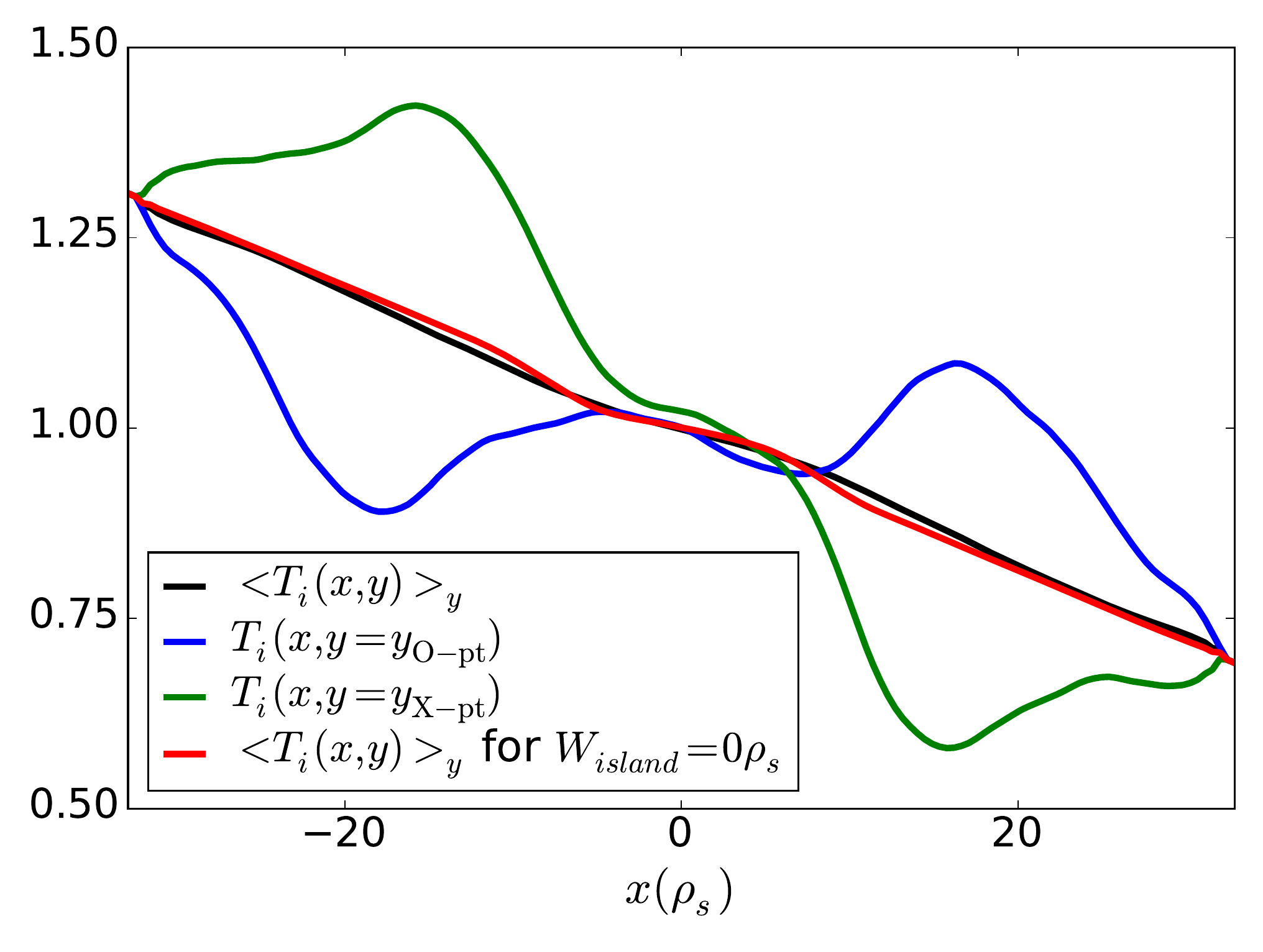}
\caption{Profiles of the total ion temperature $T_i=T_{i0}+\tilde{T}_i$. The black curve is the poloidal average, the blue (respect. green) curve is the profile across the O-point at $y=0\rho_s$ (respectively X-point at $y=\pm 32\rho_s$) with $W_{island}=15\rho_s$. Red curve is the poloidal average of the total ion temperature profile with no island $W_{island}=0$.}
\label{fig.Ti_1D}
\end{figure}

In Fig.~\ref{fig.Esat.vs.Wisland} the saturated energy $E_{\rm sat}$ averaged on the time range $t \in \left[ 500; 3000 \right] a / c_s$ is plotted against the size of the static magnetic island as well as its standard deviation. We observe that beyond a threshold of the island size (around $5 \rho_s$) the saturated energy increases significantly. Further analysis has shown that it is due to an increase of the ion temperature fluctuations and not to the island magnetic energy. Indeed, as mentioned previously, the dominant part of the energy given by~\refeq{eq.E} is the fluctuations of the ion temperature $1/2 \int \tau \tilde{T}_i^2 / (\Gamma-1)$. The values of $\tilde{T}_i$ fluctuate around $0.4$ with no island and around $0.7$ with $W_{island}=15\rho_s$. The amplitude of the fluctuations are approximately $2$ times bigger (i.e., $\sim \pm 0.2$) for $W_{island}=15\rho_s$ with respect to the one with no island. Fig.~\ref{fig.Ti_2D} shows time slices at times $t=\{1506,1516,1526,1536,1546\} a/c_s$ of the ion temperature $T_{i0}+\tilde{T}_i$ without ($W_{island}=0 \rho_s$) and with ($W_{island}=15 \rho_s$) a static magnetic island. At this particular time slices with a static magnetic island, Fig.~\ref{fig.Ti_2D.b}, \ref{fig.Ti_2D.d}, \ref{fig.Ti_2D.f}, \ref{fig.Ti_2D.h} and \ref{fig.Ti_2D.j} show the propagation of an ion temperature finger-like structure across the island associated with a clockwise flow inside the island. This propagation is due to the enhancement of the perpendicular heat flux transport close to the X-point and the clockwise flow rotation inside the island. At other time slices, we can observe opposite characteristics with a radial symmetry with respect to $x=0$ for lower ion temperature structures associated with counter-clockwise flow rotation inside the island. Fig~\ref{fig.Ti_2D_avg} shows the time average of the ion temperature over $t\in [500,3000]\ a/c_s$. We observe the expected flattening of the ion temperature due to the effect of the island. However, this flattening is not poloidally invariant. The flattening seems to be extended toward the top of the island. This is due to the direction of the poloidal zonal flow which goes along decreasing poloidal coordinates at the middle of the box (i.e., $\vert x \vert < 10$) and along increasing poloidal coordinates at the edge of the box (i.e., $\vert x \vert > 10$). Moreover, an additional poloidal variation of the flattening is observed in Fig.~\ref{fig.Ti_1D} which contains the poloidal average $<T_{i0}+\tilde{T}_i>_{y,t}$ of the total ion temperature as well as the profiles across the X-point $<T_{i0}+\tilde{T}_i(y=b/2)>_t$ and the O-point $<T_{i0}+\tilde{T}_i(y=0)>_t$.
As a result, the flattening of the ion temperature is dominant on the profile across the O-point ($y=0$). As explained with the previous figure, this is due to the fact that all positive fluctuations of the ion temperature (i.e. a finger-like structure of ion temperature) which appear close to the X-point enter inside the island from the left hand side, follow the magnetic flux surfaces toward the right hand side where this finger diffuses and can cross the separatrix toward the right hand side. The clockwise rotation of the flow inside the island is due to the positive radial transport (toward increasing radial position) for positive fluctuations of the ion temperature. In contrast, when negative fluctuations of the ion temperature appear close to the X-point, the counter-clockwise rotation dominates due to the negative radial transport (toward decreasing radial position) of negative fluctuations of the ion temperature.

Finally, due to our forcing of the conservation on average of the background ion temperature gradient by the presence of the viscosity term $\nu_s$ in the equations, the flattening of the total ion temperature across the O-point produces an increase of the ion temperature gradient around the X-point. Moreover, by looking at smaller variations of the time averaged total ion temperature profile across the O-point we can see that around the positions of the inner (at $x=-15\rho_s$) and outer (at $x=15\rho_s$) separatrix, the gradients are inverted and become locally positive. The effect of these inversions of the ion temperature gradient is to generate local variation of the flow. We remark that close to the middle position $x=0\rho_s$, either with or without the background static island, a very small flattening tendency of the ion temperature profile appears due to a small magnetic shear. This flattening is negligible in comparison to the flattening due to the magnetic island.

\begin{figure}[!htbp]
\centering
\subfigure[\ Radial heat flux for $W_{island}=0 \rho_s$]{
\includegraphics[width=79mm]{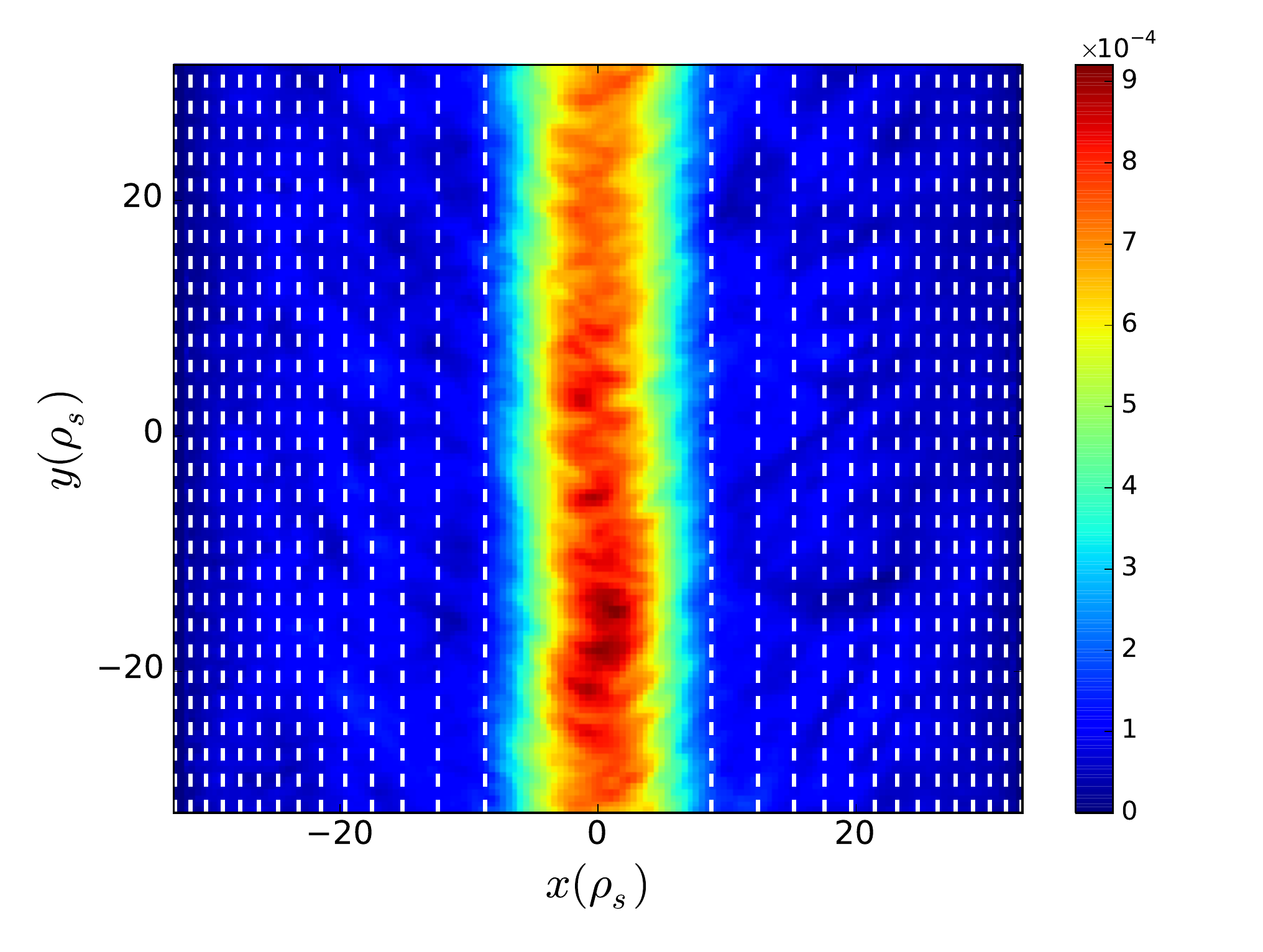}
}
\subfigure[\ Radial heat flux for $W_{island}=10 \rho_s$]{
\includegraphics[width=79mm]{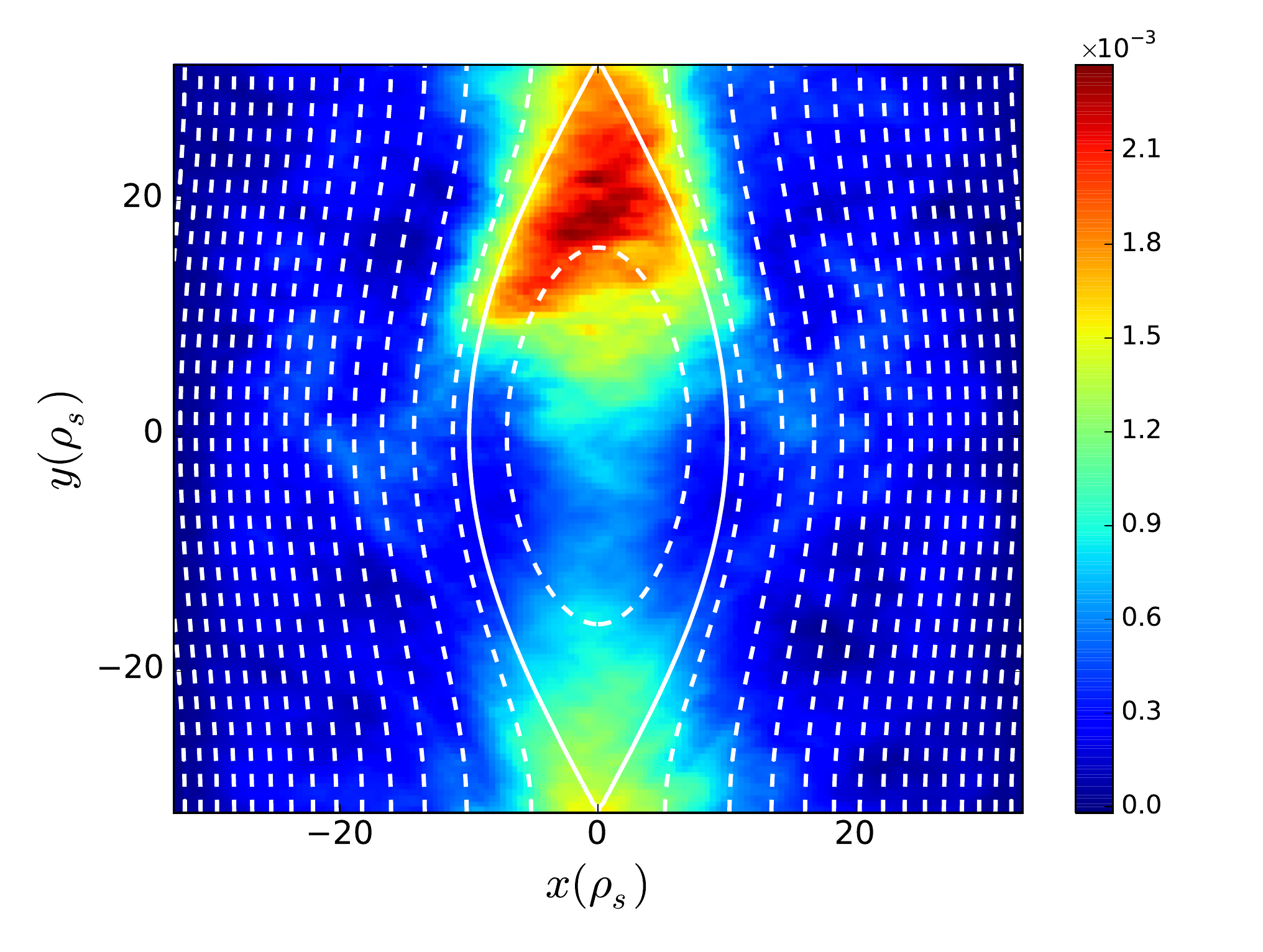}
}
\subfigure[\ Radial heat flux for $W_{island}=20 \rho_s$]{
\includegraphics[width=79mm]{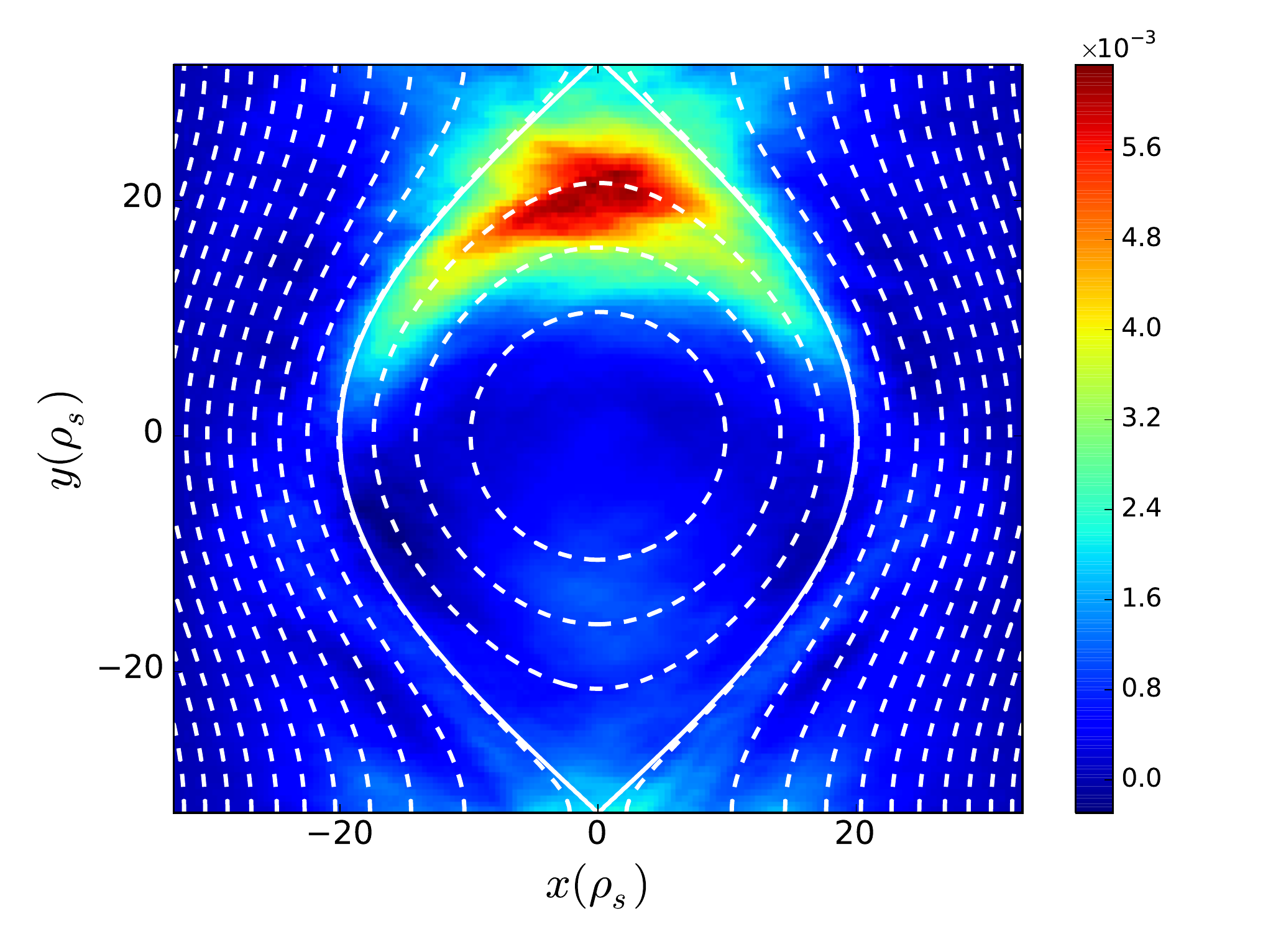}
}
\caption{Comparison of the time average of the radial heat flux $Q = \tilde{v}_x \tilde{T}_i$ over the time range $t \in \left[ 500, 3000 \right]\ a / c_s$ between the following sizes of the static magnetic island $W_{island} \in \{ 0; 10; 20 \} \rho_s$. The white plain (resp. dashed) curves are the island separatrix (resp. 15 iso-contour magnetic flux surfaces). The radial ion heat flux is peaked inside the island between the X-point and the O-point.}
\label{fig.Flux2D.vs.Wisland}
\end{figure}
In Fig.~\ref{fig.Flux2D.vs.Wisland} we can see the radial ion heat flux $<\tilde{v}_x \tilde{T}_i>$ due to the $\bm{E} \times \bm{B}$ drift averaged over the time range $t \in \left[ 500; 3000 \right]\ a / c_s$ for different sizes of the island $W_{island} \in \{ 0; 10; 20 \} \rho_s$ respectively for the figures (a), (b) and (c). With no static island (a), the flux is concentrated around the resonant surface at $x=0$. This time averaged radial ion heat flux is almost homogeneous in the poloidal direction. By increasing the size of the static magnetic island, we observe a compression of the time averaged ion heat flux toward the X-point but, more precisely toward the top of the island. Indeed, the poloidal profile is more peaked near $y \sim 20 \rho_s$ and the maximum value of this poloidal profile is approximately $3$  times higher for $W_{island} = 20 \rho_s$ than for $W_{island} = 10 \rho_s$ and approximately $10$ times higher than for the case with no island. We observe that the shape of the radial ion heat flux is consistent with the previous observation of the flattening of the ion temperature across the O-point. We also remark that the radial ion heat flux average is positive because the radial transport is such that $\tilde{v}_x$ is negative (respectively positive) for negative (respectively positive) ion temperature fluctuations $\tilde{T}_i$. To obtain more detailed understanding of the global effects of the island on the turbulence and transport, we need to compare the radial ion heat flux by averaging Fig~\ref{fig.Flux2D.vs.Wisland}, over the poloidal dimension $y$.\\
%
%
\begin{figure}[!ht]
\centering
\subfigure[\ Width $W_Q$ of the fitting Gaussian profile]{
\includegraphics[height=30mm]{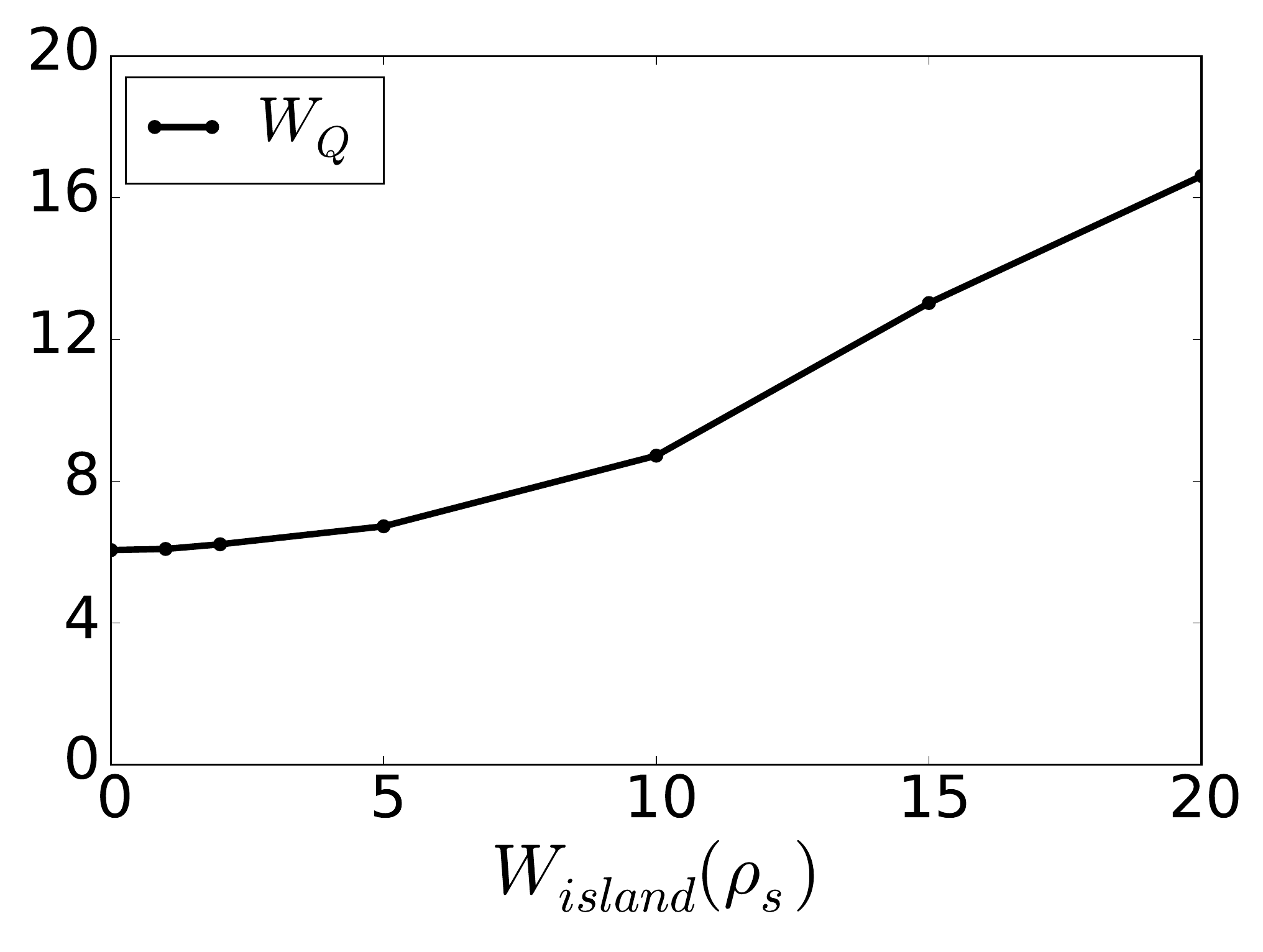}
}
\subfigure[\ Max $Q_0+Q_1$ and min $Q_0$ of the fitting Gaussian profile]{
\includegraphics[height=30mm]{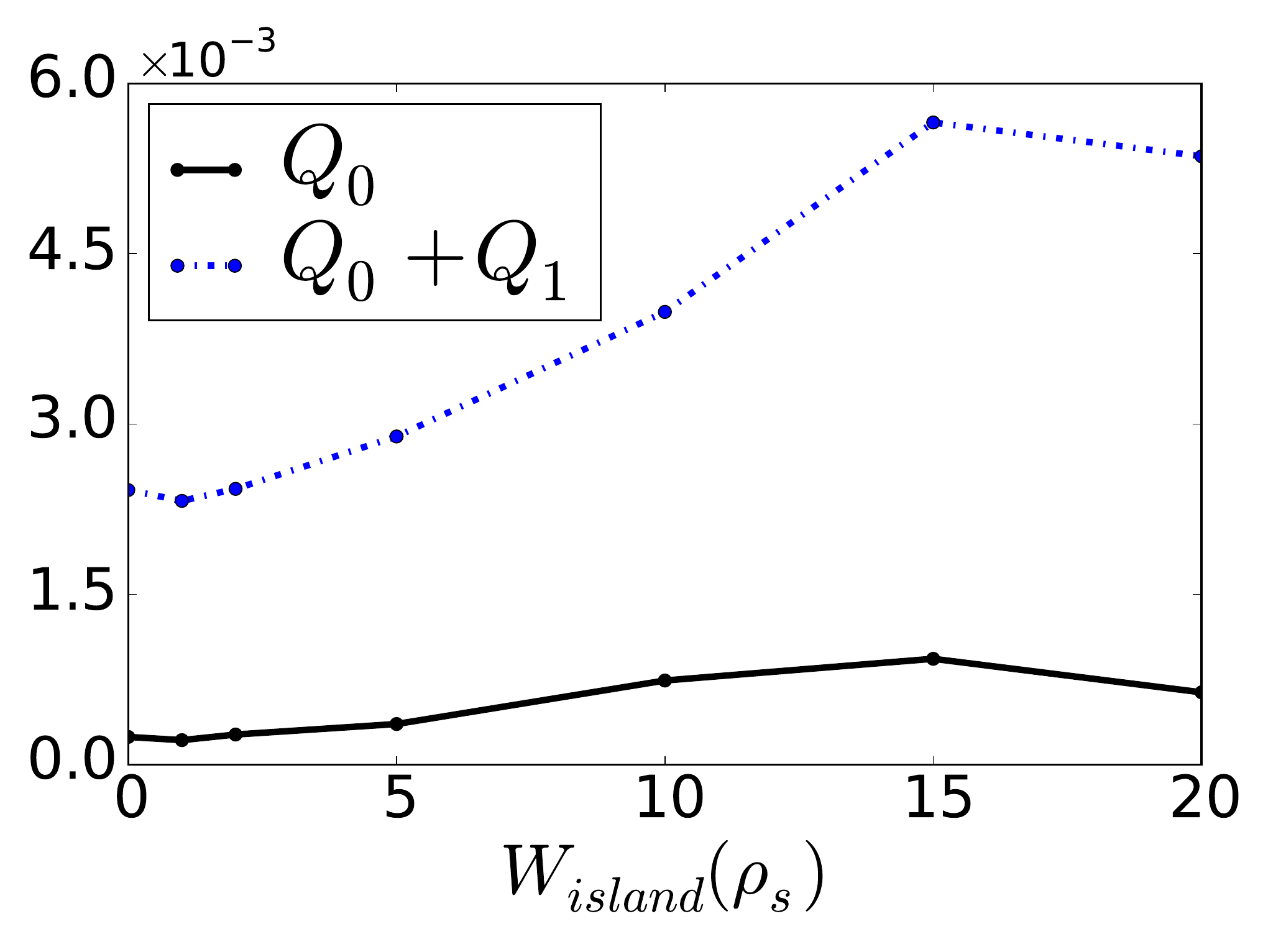}
}
\subfigure[\ Width $W_{\phi}$ of the fitting Gaussian profile]{
\includegraphics[height=30mm]{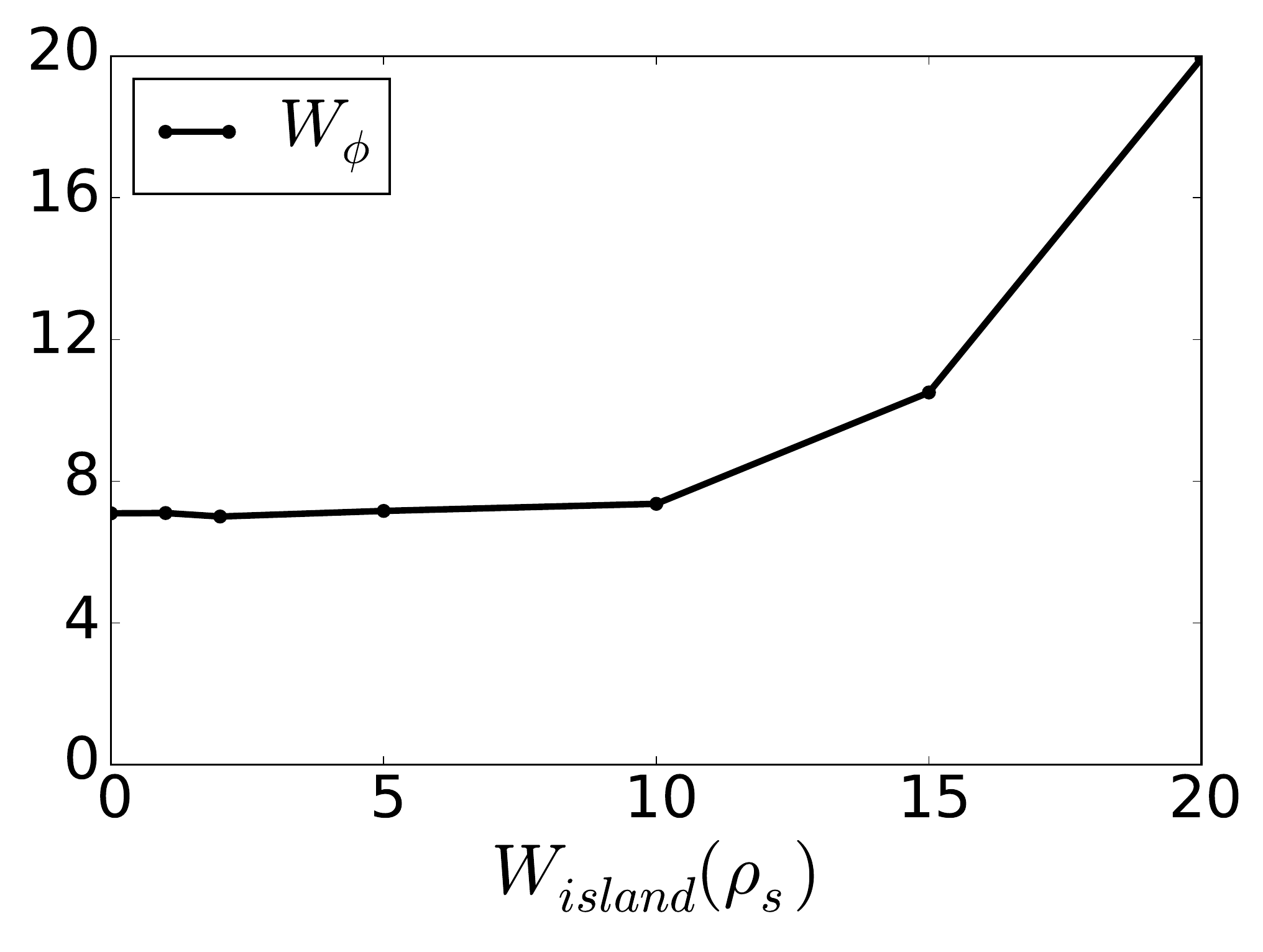}
}
\subfigure[\ Max $\phi_0+\phi_1$ and min $\phi_0$ of the fitting Gaussian profile]{
\includegraphics[height=30mm]{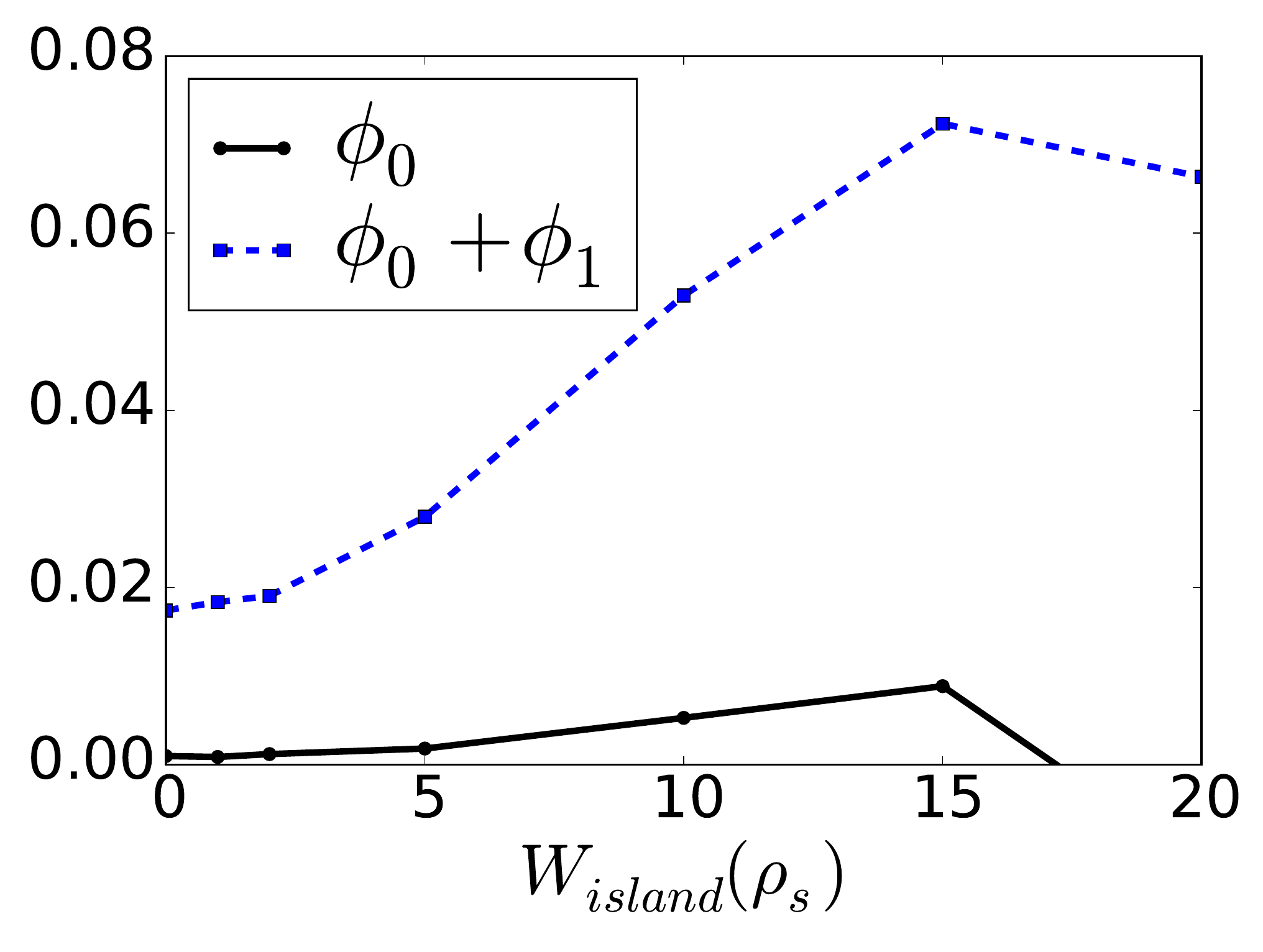}
}
\caption{Comparison of the Gaussian fitting parameters against the size of the static magnetic island $W_{island} \in \{ 0; 1; 2; 5; 10; 15; 20 \} \rho_s$ for (a)-(c) the radial ion heat flux and for (b)-(d) the square of the electrostatic potential. The time range is $t \in \left[ 500, 3000 \right] a/c_s$. Beyond a threshold around $W_{island} \sim 5\rho_s$, the static island globally enhances the transport and turbulence even outside the island. 
}
\label{fig.Flux2D.Parameterization}
\end{figure}
We use a parameterization of the poloidal average $\left< \cdots \right>_{y}$ of the time-averaged ion heat fluxes of Fig.~\ref{fig.Flux2D.vs.Wisland}. The parameterization is done using a fitting of the profiles with the following Gaussian
\bgeqa
\displaystyle
f(x) = f_0 + f_1 \exp\left(-\frac{x^2}{W_f^2} \right).
\edeqa
The variation of the parameter $f_0$ describes the effects of the island outside the island (i.e., $\vert x \vert > W_{island}$), $f_0+f_1$ the effects around the position $x=0$ inside the island, and $W_f$ the characteristic width of the Gaussian profile. Fig.~\ref{fig.Flux2D.Parameterization} represents the variations of these parameters used by the Gaussian fitting of the radial ion heat flux $Q = \tilde{v}_x \tilde{T}_i$ corresponding to the transport property and the squared electric potential $\tilde{\phi}^2$ corresponding to the turbulence property.

We observe the fact that the width of the Gaussian profile fits as well as the maximum value at $x=0$ increase for the transport and the turbulence beyond a threshold island size. The threshold is around $5\rho_s$ for the transport and between $5\rho_s$ and $10\rho_s$ for the turbulence. These thresholds are about the same order as the characteristic length of the turbulence $5\rho_s$ to $10 \rho_s$ of our simulations. Moreover, the parameters $Q_0$ and $\phi_0$ correspond to the effect of the island respectively on the transport and the turbulence outside the island. Even if it is a small variation, we observe an increase of the transport and the turbulence far away from the island. This observation means that there is an enhancement of the turbulence and the transport independent of a simple displacement by the presence of the island.

\begin{figure}[!htps]
\centering
\subfigure[\ Flux surface average of the heat flux inside the island.]{
\label{fig.Flux.FluxSurfAvg.a}
\includegraphics[width=85mm]{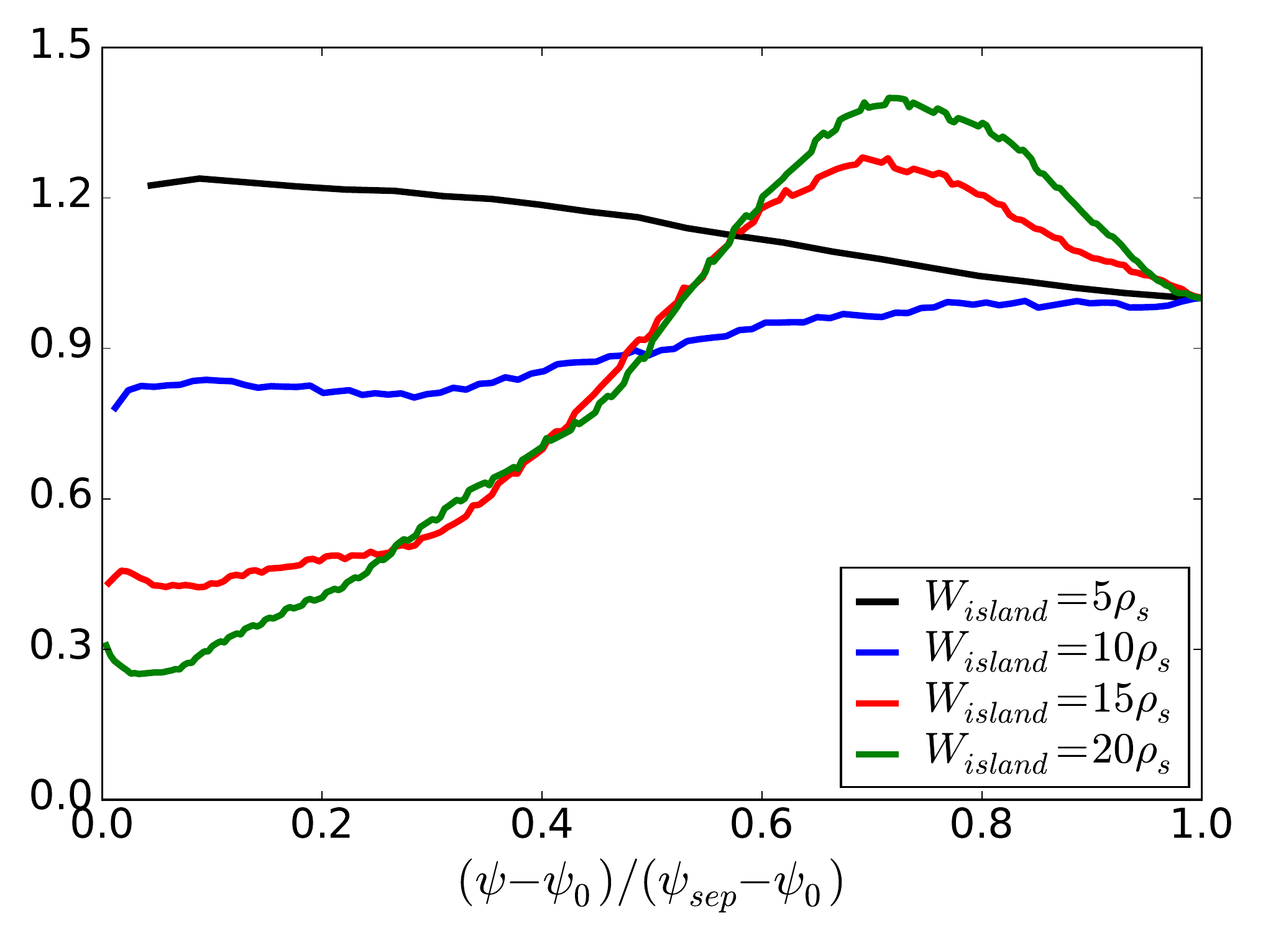}
}
\subfigure[\ Flux surface average of the heat flux outside the island.]{
\label{fig.Flux.FluxSurfAvg.b}
\includegraphics[width=85mm]{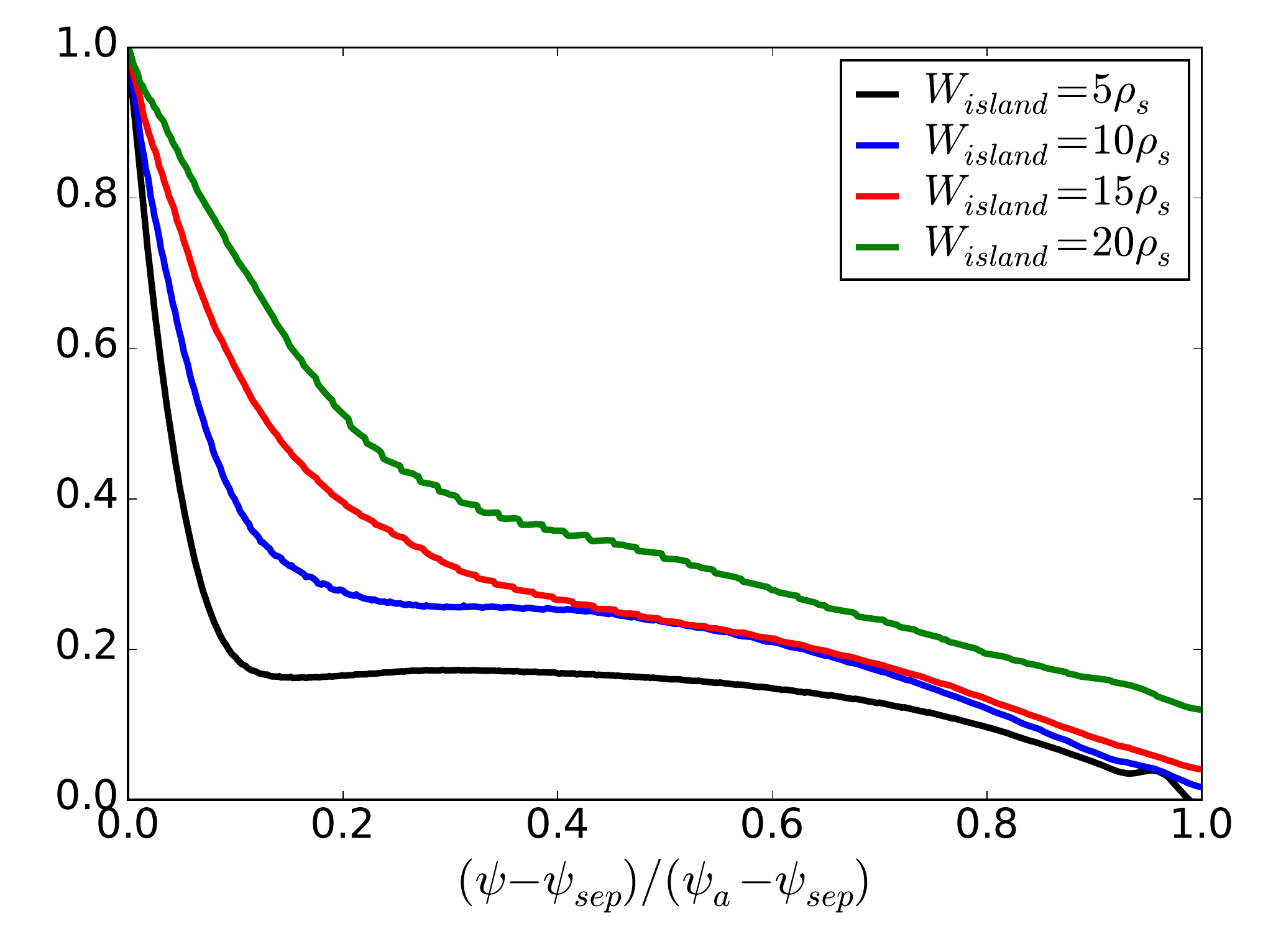}
}
\caption{Comparison of the time and flux surface averages of the radial ion heat flux for the following sizes of the static magnetic island $W_{island} \in \{ 5; 10; 15; 20 \} \rho_s$. The effect of the poloidal asymmetry inside the island and the threshold of $W_{island} \sim 5\rho_s$ are consistent with these flux surface averages.}
\label{fig.Flux.FluxSurfAvg}
\end{figure}

Finally, these results given by a simple poloidal average can be verified by using another average which conserves the structure of the static magnetic island. 
To meet this goal, a last investigation is to try to recover the previous results by using the flux surface average $\left< \cdots \right>_{\psi_{eq}}$ instead of the poloidal average $\left< \cdots \right>_{y}$. Fig.~\ref{fig.Flux.FluxSurfAvg} represents the flux surface averages of the time-averaged radial ion heat flux for different sizes of the magnetic island. The first figure corresponds to profiles of the radial ion heat flux inside the island versus the flux surface coordinate $(\psi-\psi_{0})/(\psi_{sep}-\psi_{0})$ for different sizes of the magnetic island and the second figure to profiles outside the island versus $(\psi-\psi_{sep})/(\psi_{a}-\psi_{sep})$. The values $(\psi_0,\psi_{sep},\psi_a)$ of the magnetic flux correspond respectively to the O-point, the separatrix, and the last closed flux surface in our simulation box. Beyond the last closed flux surface, all integrals along each opened flux surface linearly decrease to $0$, which is their minimum values at the corner of the simulation box of the flux surface corresponding to $\psi_{eq}(-a,b/2)$. As a summary, for Fig.~\ref{fig.Flux.FluxSurfAvg.a} (respectively Fig.~\ref{fig.Flux.FluxSurfAvg.b}) corresponding to the radial ion heat flux inside (respectively outside) the island, the abscissa coordinate represents the O-point (respectively the separatrix) at the value $0$ and the separatrix (respectively the last closed flux surface) at the value $1$.
We observe inside the island (Fig.~\ref{fig.Flux.FluxSurfAvg.a}) an increase of the shift of the radial ion heat flux toward the separatrix instead of being homogeneous everywhere inside the island. However, the asymmetry between the top and the bottom of the island cannot be recovered in comparison to what have been shown above with the 2D slices. With the observation of the radial ion heat flux outside the island (Fig.~\ref{fig.Flux.FluxSurfAvg.b}), the background transport increases beyond a threshold size of the magnetic island between $5 \rho_s$ and $10 \rho_s$, related to the characteristic length of the transport with no island. This enhancement of the transport outside the island is due to the filaments which appear at different time slices as described previously. As an example with a static magnetic island $W_{island}=20\rho_s$, we typically observe filaments of the positive fluctuations of the ion temperature starting around the position $(x,y)=(-10,25)$, crossing the separatrix, following the magnetic flux surfaces until the position $(x,y)=(20,0)$ and finally diffusing outside the island in the region of lower ion temperature.

\begin{figure}[!htps]
\centering
\subfigure[\ $\left< \phi \right>_t$ for $W_{island}=0\rho_s$.]{
\label{fig.kyMODES.all.W0}
\includegraphics[width=40mm]{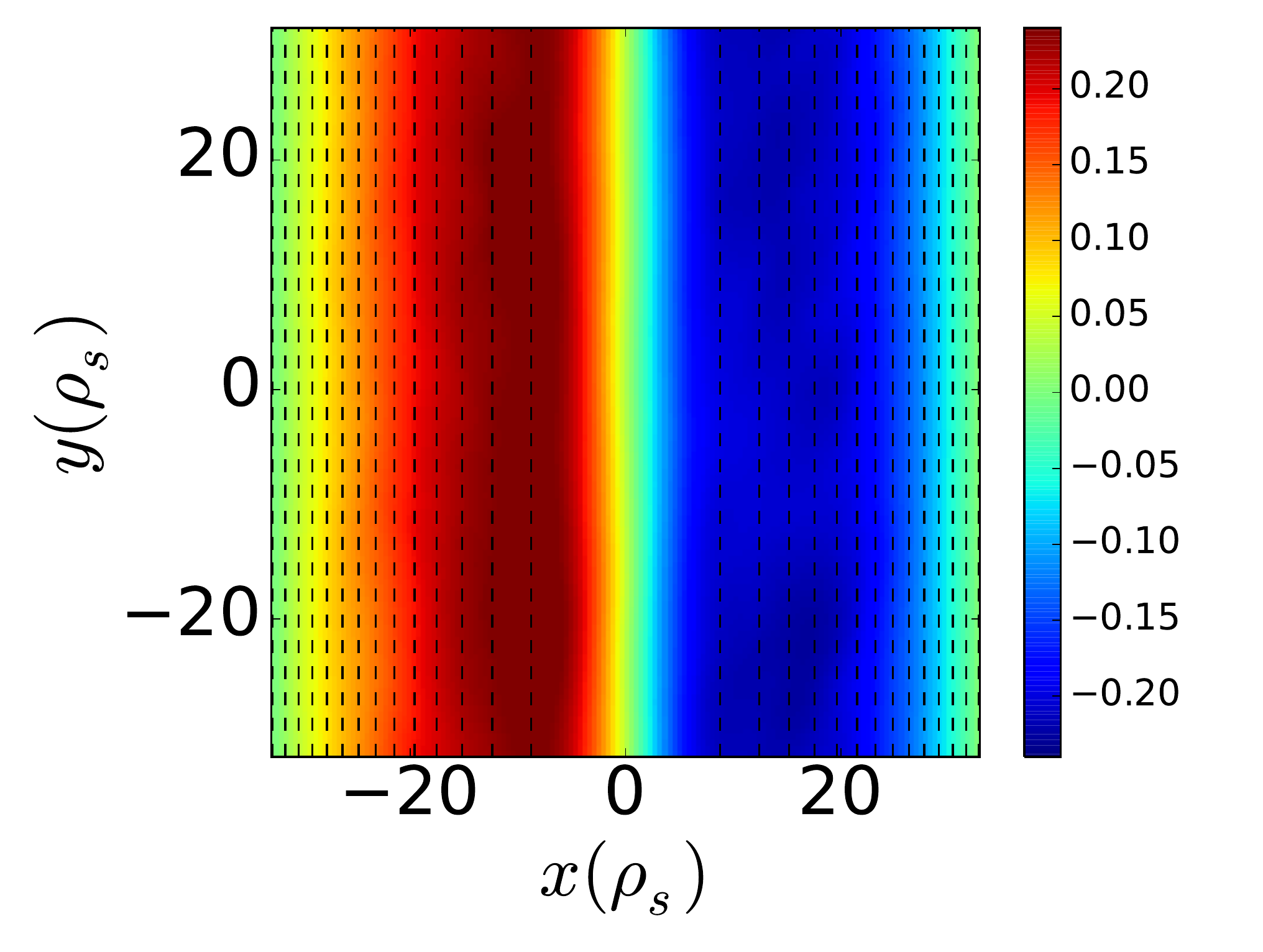}
}
\subfigure[\ $\left< \phi \right>_t$ for $W_{island}=15\rho_s$.]{
\label{fig.kyMODES.all.W15}
\includegraphics[width=40mm]{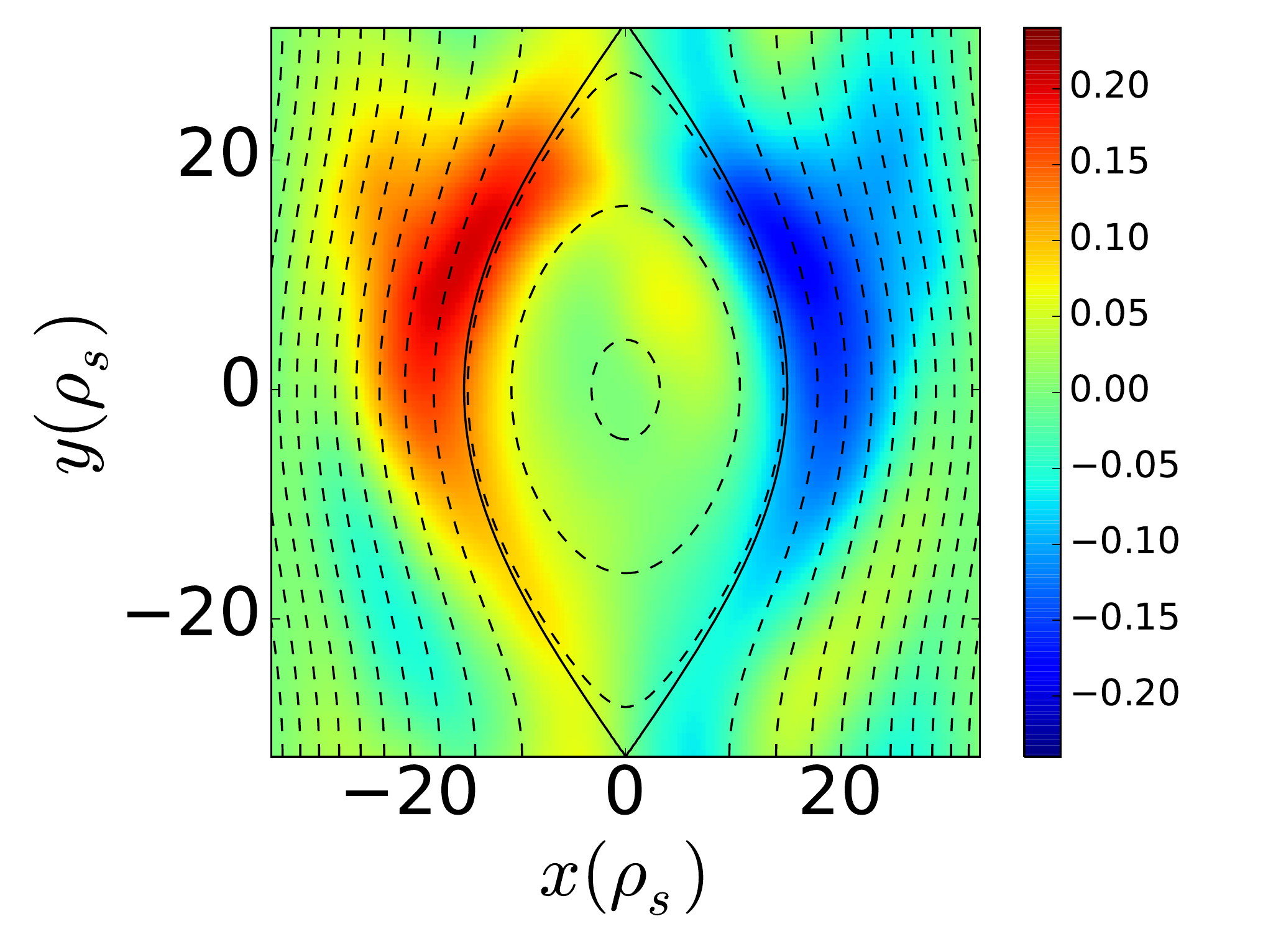}
}
\caption{Comparison of the time averaged electric potential $\left< \phi \right>_t$ on $t \in \left[ 500; 3000 \right]\ a/c_s$ for a static island of width (a) $W_{island}=0\rho_s$ and (b) $W_{island}=15\rho_s$.}
\label{fig.kyMODES.all}
\end{figure}
\begin{figure}[!htps]
\centering
\subfigure[\ Mode $k_y=0$ for $W_{island}=0\rho_s$.]{
\label{fig.kyMODES.0.W0}
\includegraphics[width=40mm]{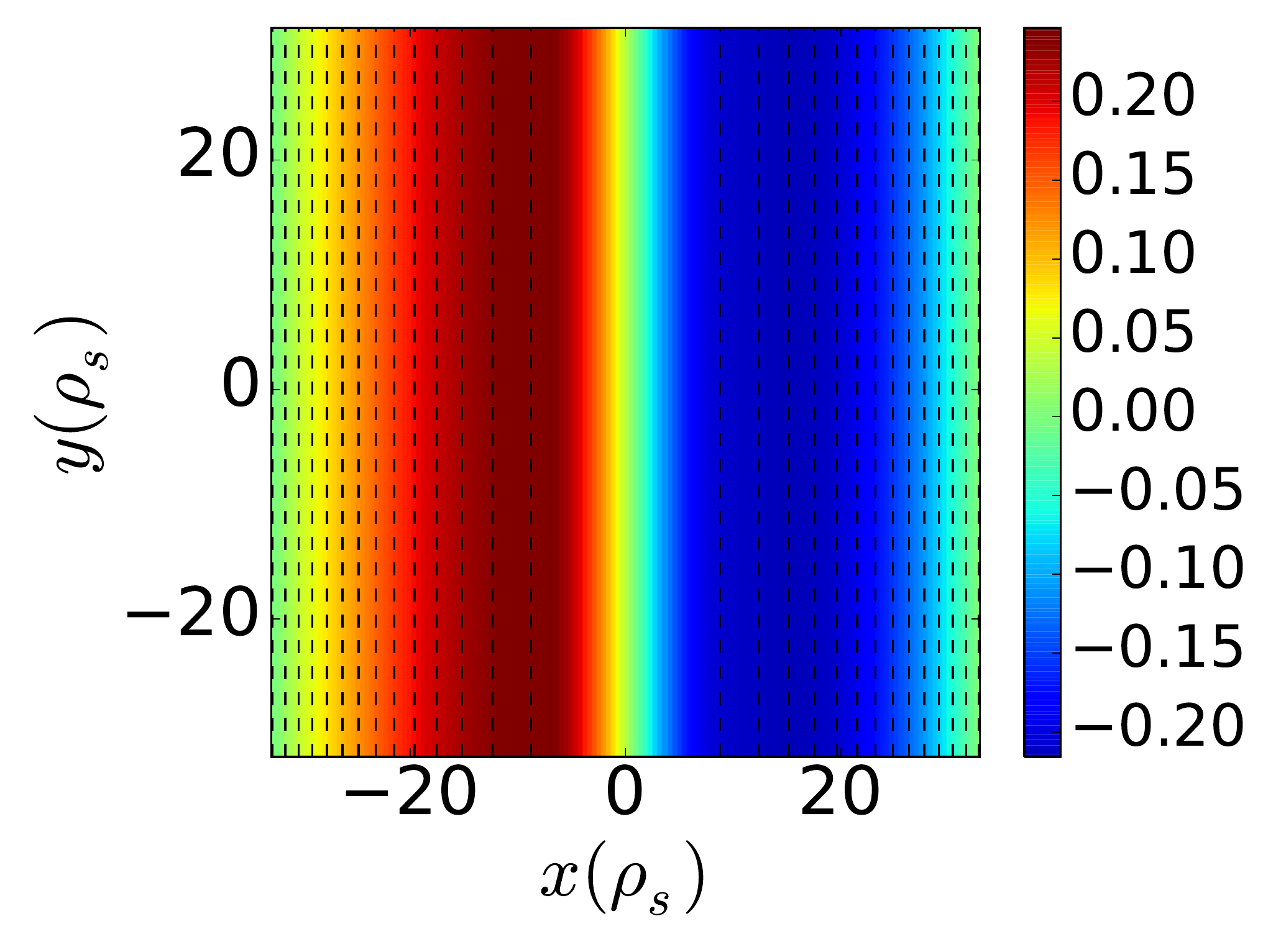}
}
\subfigure[\ Mode $k_y=0$ for $W_{island}=15\rho_s$.]{
\label{fig.kyMODES.0.W15}
\includegraphics[width=40mm]{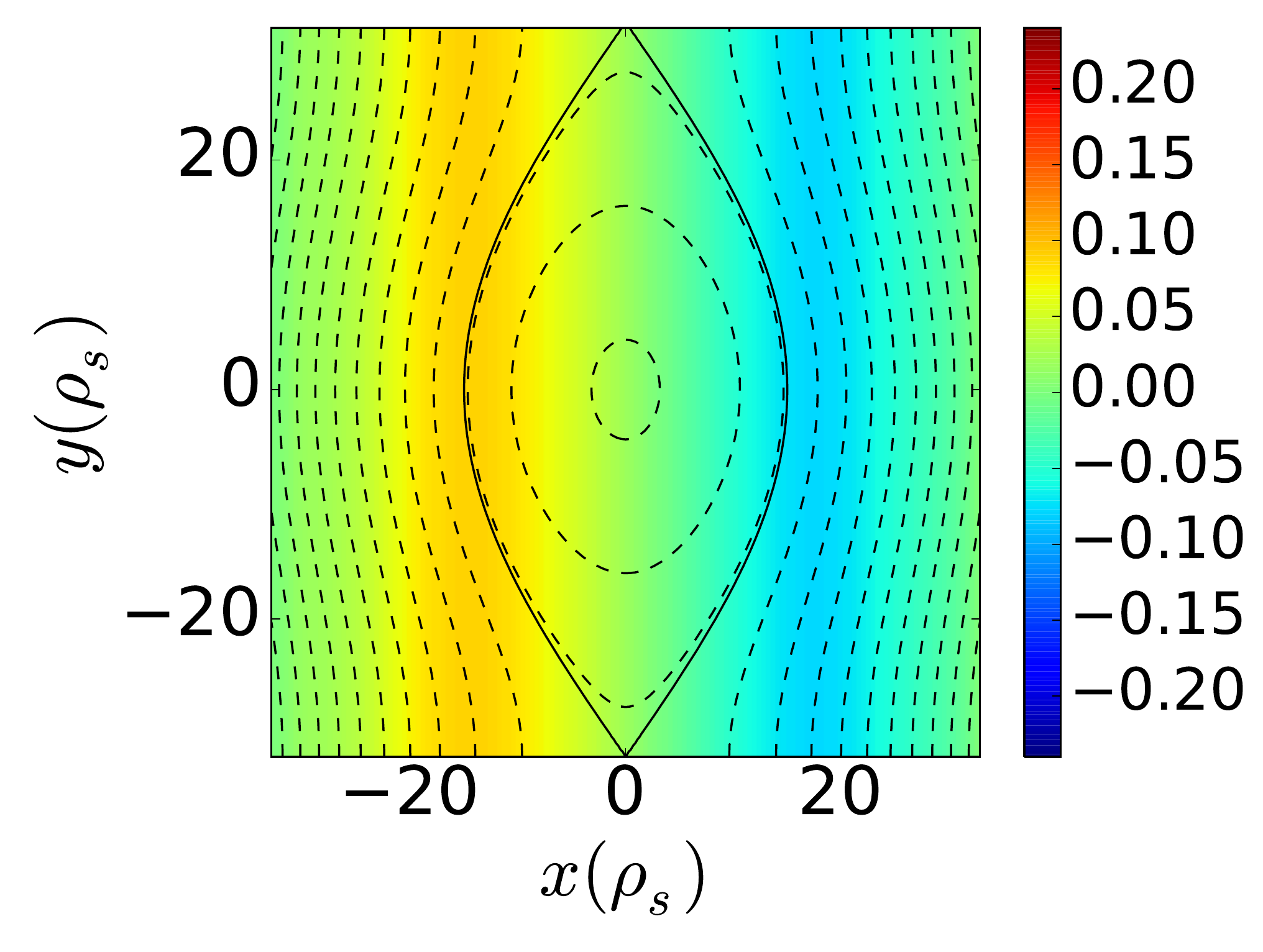}
}
\subfigure[\ Mode $k_y=\pm 2\pi/b$ for $W_{island}=0\rho_s$.]{
\label{fig.kyMODES.pm1.W0}
\includegraphics[width=40mm]{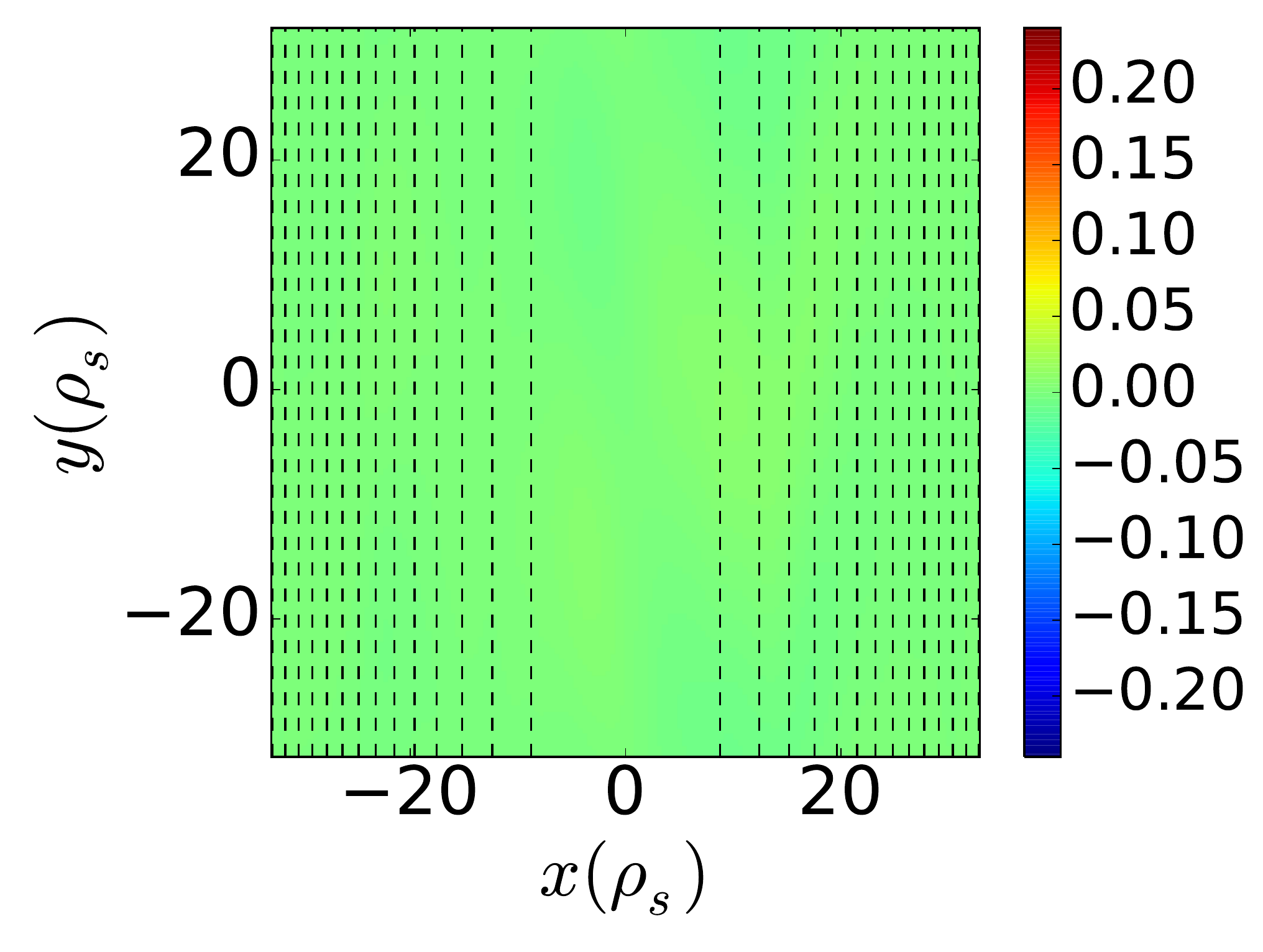}
}
\subfigure[\ Mode $k_y=\pm 2\pi/b$ for $W_{island}=15\rho_s$.]{
\label{fig.kyMODES.pm1.W15}
\includegraphics[width=40mm]{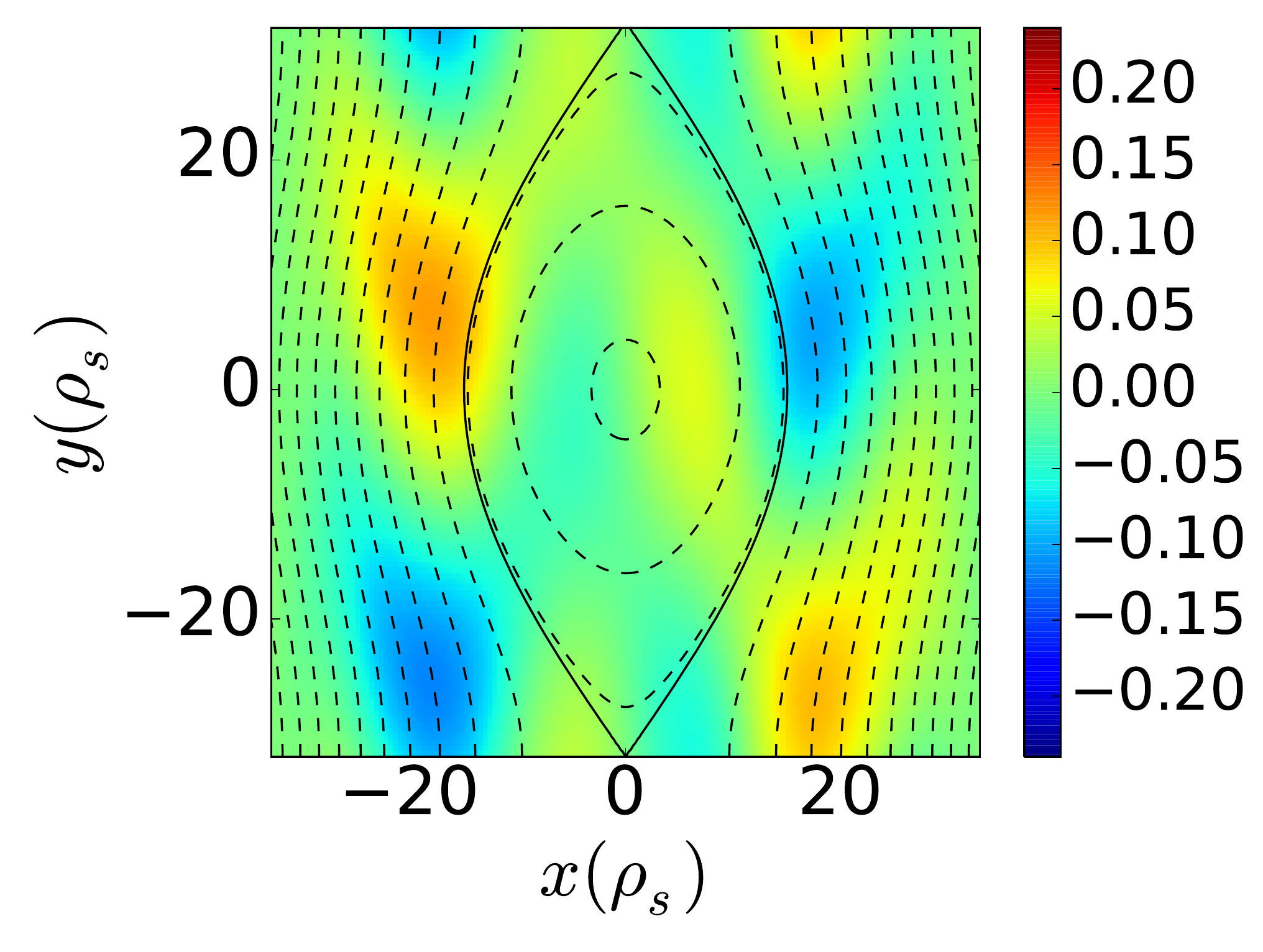}
}
\caption{Reconstruction of the 2D electric potential from one selected poloidal mode: (a)-(b) from $k_y=0$ and (c)-(d) from $k_y=\pm 2\pi/b$ with (a)-(c) no island and (b)-(d) $W_{island}=15\rho_s$.}
\label{fig.kyMODES.2D}
\end{figure}
It is important to note that in addition to the turbulent heat flux, there is also transport due to static $\bm{E}\times\bm{B}$ flows which arise due to the presence of the island, that will also drive transport.  Fig.~\ref{fig.kyMODES.all} shows the time-averaged electrostatic potential for the cases of $W_{island} = 0\rho_s$ and $15 \rho_s$, respectively.  In the no-island case, the mean potential corresponds to a $k_y = 0$ zonal flow with radial wavenumber approximately equal to $\pi/a$.  However, in the finite island case (Fig.~\ref{fig.kyMODES.all.W15}), the mean potential has both significant $k_y=0$ and $k_y=2\pi/b$ components, with comparable magnitude to the no-island case.  These components are separately visualized in the (x,y) plane in Fig.~\ref{fig.kyMODES.2D}.  Note in particular that the finite $k_y$ equilibrium flow in the finite island case is not symmetric about the O-point at $y=0$, and thus contributes to the asymmetry in radial heat flux shown in Fig.~\ref{fig.Flux2D.vs.Wisland}.

As a conclusion of our investigation of the effects of a static island on turbulence and transport, we observed the flattening of the ion temperature profile across the O-point ($y=0$) due to the island as well as an increase of its gradient across the X-point. The increase of the ion temperature gradient around the X-point enhances the ITG instability and drives more turbulence in this region. Due to the radial transport, positive or negative structures of the ion temperature enter inside the island and flow along the magnetic flux surfaces. Inside the island, the time average of the radial ion heat flux is then peaked on the top of the island due to the fact that the flow naturally goes along decreasing poloidal coordinates around the radial position $x=0$ (as observed with no island). This privileged direction of the flow is due to our choice of the sign of the equilibrium poloidal magnetic flux (i.e., the sign of the magnetic shear). Finally, because of the presence of the island, the turbulence and transport are significantly increased when the size of the static magnetic island is bigger than the characteristic size of the turbulence and transport.

To make progress toward understanding the self-consistent interaction, some results of the feedback of the modified turbulence on the dynamical evolution of the island is investigated below.

\section{First results toward the feedback on the island}
\label{ToC.EtaTurb}
One of our interests is to understand the self-consistent interaction between fast dynamics of microturbulence and the slow dynamics of a magnetic island. In the previous section we focused on the effect of a static magnetic island on transport and ITG microturbulence. From these results, we report here a quantitative study of the effect of microturbulence on the slow dynamics of the magnetic island.

\subsection{Formulation of a turbulent resistivity}
We begin by adopting a mean-field type approach to exploit the separation of time and spatial scales between the turbulence and the island.  We can then rewrite the magnetic flux evolution equation as
\bgeqa
\label{eq.dpsiIdt}
\displaystyle
\beta \partial_t \psi_I &=& - {B}_{y,0} \partial_y \phi_I - \eta_{\rm cl} j_{z,I} + \beta \left[ \tilde{\psi} , \tilde{\phi}-\tilde{n} \right],
\edeqa
where ${B}_{y,0} \partial_y = \nabla_{\parallel}^0 = \nabla_{\parallel} - \tilde{\nabla}_{\parallel}$, the second term is the usual classical resistive current and the last term corresponds to the average of the Ohm's law stress. This last term can be written as a coherent part $- \eta_{\rm turb} j_{z,I}$ and a non-coherent part $- \eta_{\rm NC} j_{\rm NC}$ with respect to the island current where the effective turbulent resistivity describes the self-consistent feedback of the turbulence on the dynamics of the island
\bgeqa
\label{eq.etaTurb}
\displaystyle
\eta_{\rm turb} &=& - \frac{ \left< \left[ \tilde{\psi} , \tilde{\phi}-\tilde{n} \right] j_{z,I} \right>_{x,y} }{ \left< j_{z,I}^2 \right>_{x,y} }.
\edeqa
Moreover, in addition to the dynamical evolution of the magnetic flux of the island $\psi_I$, a dynamical evolution of the electric potential $\phi_I$ associated with the magnetic island has also been developed. Here, we do not focus on this part because in our previous simulations we imposed only a poloidal magnetic flux and we omit the presence of an associated electric potential. Indeed, the associated electric potential $\phi_I$ of the island is a very slow dynamical quantity in comparison to the fast evolution of $\tilde{\phi}$ and $\phi_I$ has almost no effect on the dynamical evolution of the magnetic island poloidal flux.

\subsection{Evaluation of turbulent resistivity in simulation}
For the dynamical equation of the island, simulations can be challenging since we need to resolve the fast time scales of the microturbulence as well as the slow time scales of the evolution of the magnetic island and the tearing mode. 
In this work, we have exploited the timescale separation between the island and turbulence dynamics to approximate the island as fixed, consistent with previous studies by other groups.  However, we can still begin to probe how we might expect the turbulence to act on the island in a fully dynamic model by calculating the time-averaged forcings of the turbulence on the imposed island structure.

\begin{figure}[!htbps]
\centering
\includegraphics[width=90mm]{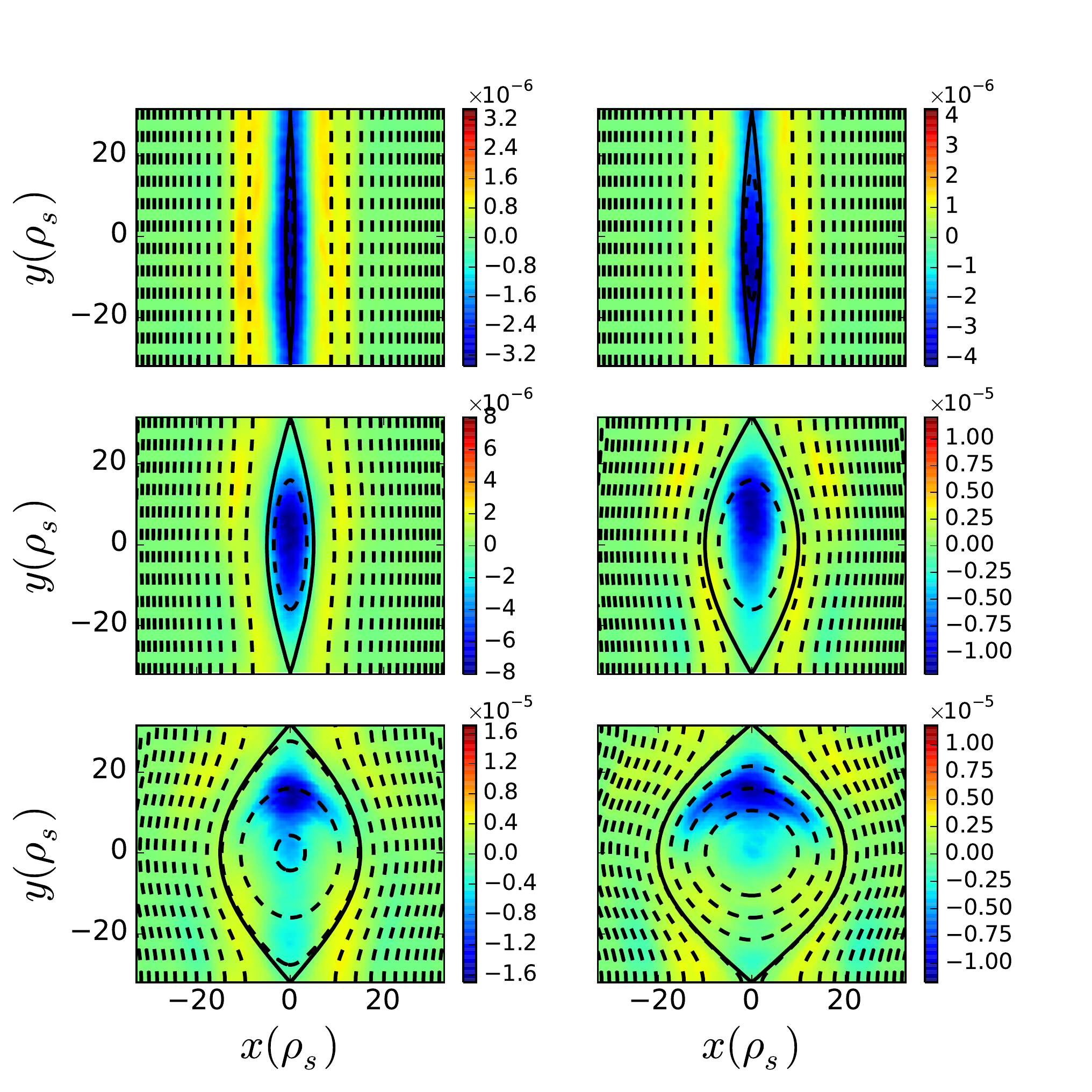}
\caption{Ohm's law stresses $\left[ \tilde{\psi} , \tilde{\phi}-\tilde{n} \right]$ averaged over $t \in \left[ 2250 ; 3000 \right]\ a / c_s$ for different sizes of the static magnetic island $W_{island} \in \{ 1; 2; 5; 10; 15; 20 \} \rho_s$ (from left to right and top to bottom). The black plain (resp. dashed) curves are the island separatrix (resp. 15 iso-contour magnetic flux surfaces). The structures of the Ohm's law stresses are consistent with the radial ion heat fluxes.}
\label{fig.OhmLawStresses}
\end{figure}
Fig.~\ref{fig.OhmLawStresses} represents the Ohm's Law stresses $\left[ \tilde{\psi} , \tilde{\phi}-\tilde{n} \right]$ averaged over $t \in \left[ 500 ; 3000 \right]\ a / c_s$ for different sizes of the static magnetic island $W_{island} \in \{ 1; 2; 5; 10; 15; 20 \} \rho_s$. There is an obvious link between the structures of these Ohm's Law stresses and those of the radial ion heat fluxes shown in Fig.~\ref{fig.Flux2D.vs.Wisland}.
\begin{figure}[htbps]
\centering
\includegraphics[width=85mm]{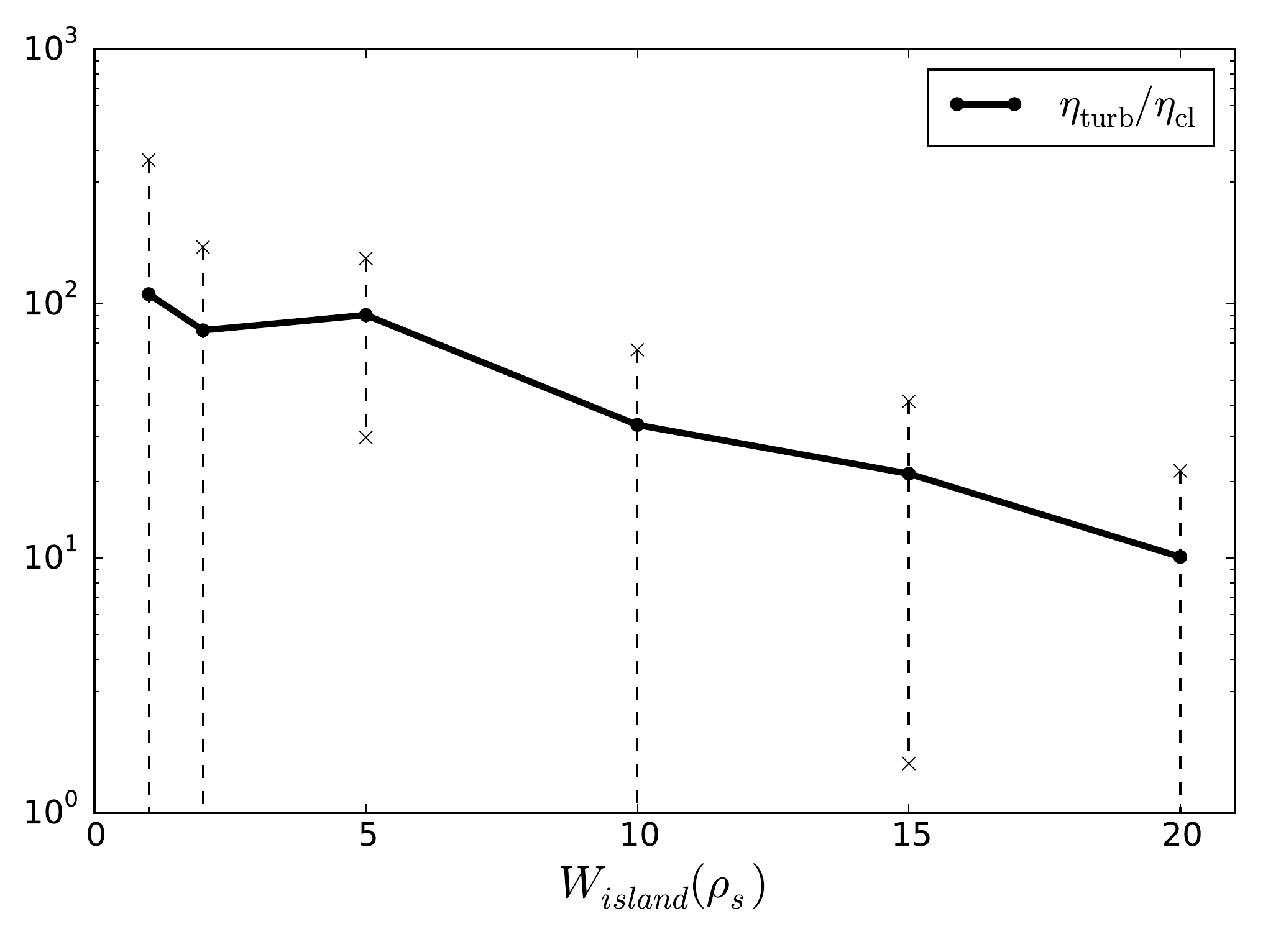}
\caption{Ratio of the effective turbulent resistivity $\eta_{\rm turb}$ given by Eq.(\ref{eq.etaTurb}) over the classic resistivity $\eta_{\rm cl}$ versus the size of the island $W_{island}$.}
\label{fig.TurbResistivity}
\end{figure}
The effective turbulent resistivity $\eta_{\rm turb}$ given by the \refeq{eq.etaTurb} is then computed from the extraction of the coherent part of the Ohm's law stress with respect to the static current associated with the magnetic island. The ratio between the effective turbulent resistivity and the classic resistivity $\eta_{\rm cl}$ is shown in Fig.~\ref{fig.TurbResistivity}. The first result is the expected fact that the effective turbulent resistivity, which quantifies the effects of the feedback of the turbulence on the dynamical evolution of the magnetic island, decreases by $1$ order of magnitude as the island size increases. The second result is the amplitude scale between the effective turbulent resistivity and the classic resistivity. Indeed, for all island sizes, the effective turbulent resistivity is at least $1$ order of magnitude higher than the classic resistivity. However, even with a difference of $2$ orders of magnitude for small islands, the total effect of the self-consistent feedback of the turbulence on the island is given by the multiplication of the effective turbulent resistivity $\eta_{\rm turb}$ with the current generated by the magnetic island $j_{z,I}$. The later is approximately $10$ times smaller for small island sizes than for larger island sizes. This investigation clarifies some expected mechanisms toward the self-consistent interaction between the ITG microturbulence and the dynamical evolution of a magnetic island. Perspectives are discussed below.

\section{Discussions and future work}
\label{ToC.Future}

In this work, we have studied the dynamics and associated thermal transport due to ITG microturbulence in the presence of static magnetic islands of varying width.  We find that the magnetic island is responsible for enhancement of the turbulent fluctuation amplitudes and radial heat flux, which increases rapidly above a threshold island width comparable to the correlation length of the turbulence in the no-island limit.  Broadly consistent with theoretical expectations, we observe a flattening of ion temperature inside the island, and localization of the radial heat flux near (but not symmetric about) the island X-point.  It should be noted that this flattening occurs due to turbulent transport processes and in-plane flow along flux surfaces, as the simulations presented here do include explicit conductive parallel heat fluxes of the form $q_{i,\parallel} = - k_{\parallel} \nabla_{\parallel} T_i$ that are often used to represent both parallel collisional dynamics and Landau-damping effects.  Thus, the combination of self-consistently calculated radial and poloidal turbulent heat fluxes in the presence of the imposed island (along with the self-consistent electrostatic flow which also forms) are sufficient to lead to temperature flattening and equilibration on the total magnetic flux surfaces.  A novel observation has been made that a asymmetric shift in location of the peak radial heat flux relative to the X-point is due to inclusion of self-consistent turbulence effects (such as the inherent phase velocity of the turbulence) which are not included in theoretical studies which simply approximate the turbulent fluxes in terms of anomalous diffusion coefficients.  In addition to calculating the response of the turbulence to the presence of the island, we also investigated the back-reaction of the turbulence on the island, using a mean-field approach to calculate an effective turbulent resistivity which would act on the island structure.  For the parameters considered, our simulations predict that while this turbulent resistivity is always at least an order of magnitude larger than the collisional value used, it is most effective at small island widths and decreases quickly with increasing island size.

There are a wide variety of future directions to be pursued in future work, building upon these results.  Foremost are inclusion of parallel heat flux due to both collisions and Landau damping, as well as fully 3D effects, to make closer connection with experiments.  Equally important in our opinion is to transition from statically imposed islands to self-consistently evolving islands driven by tearing unstable current profiles.  An important numerical challenge to be solved for these simulations in 3D is the implementation of preconditoners and implicit advance schemes suitable for long-time integration of the slowly evolving island in the presence of fast ITG turbulence and even faster damped Alfv\'{e}n waves in 3D geometry.  Greater understanding the self-consistent transport enhancements and profile flattening effects, as well as turbulent forcings of the island are needed to develop a fully predictive model of coupled tearing mode and turbulence dynamics.  We can then begin to include more realistic geometric effects in the equilibrium structure, to properly describe the coupled turbulence and reconnection processes which drive observed magnetic islands.

\section*{Acknowledgements}
This work was supported by U.S. DoE Contract No. DE-SC0007783 and DE-SC0010520. 
O.I. would like to thank Orso Meneghini for his help with OMFIT. 
Computing time was provided on the Triton Shared Computing Cluster (TSCC) at the San Diego Supercomputer Center (SDSC).


\begin{thebibliography}{AAA00}
\expandafter\ifx\csname url\endcsname\relax\def\url#1{\texttt{#1}}\fi

\bibitem{Doyle:07}
{E.J.~Doyle, et al.},
\textit{ Nucl. Fusion} \textbf{47}, S18 (2007). \url{dx.doi.org/10.1088/0029-5515/47/6/S02}. \textit{ Nucl. Fusion}
\textbf{48}, 099801 (2008). \url{dx.doi.ord/10.1088/0029-5515/48/9/099801}
\bibitem{Horton_99_RMP}
{W.~Horton},
\textit{ Rev. Mod. Phys.} \textbf{71}, 735 (1999). \url{dx.doi.org/10.1103/RevModPhys.71.735}
\bibitem{Furth:63}
{H.P.~Furth},
\textit{  Phys. Fluids} \textbf{6}, 48 (1963). \url{dx.doi.org/10.1063/1.1724507}
\bibitem{Helander:02}
{P.~Helander and D.J.~Sigmar},
\textit{ Collisional Transport in Magnetized Plasmas}, Cambridge University Press, Cambridge, UK (2002). \url{adsabs.harvard.edu/abs/2002ctmp.book.....H}
\bibitem{Freidberg:87}
{J.P.~Freidberg},
\textit{ Ideal magnetohydrodynamics}, Plenum Press, New York, NY (1987).
\bibitem{Biskamp:93}
{Biskamp},
\textit{ Phys. Fluids B} \textbf{5}, 3893 (1993). \url{dx.doi.org/10.1063/1.860612}
\bibitem{Scott:85}
{B.D.~Scott, A.B.~Hassam and J.F.~Drake},
\textit{ Phys. Fluids} \textbf{28}, 275 (1985). \url{dx.doi.org/10.1063/1.865197}
\bibitem{Fitzpatrick:95}
{R.~Fitzpatrick},
\textit{ Phys. Plasmas} \textbf{2}, 825 (1995). \url{dx.doi.org/10.1063/1.871434}
\bibitem{Diamond_ZF_review_PPCF:05}
{P.H.~Diamond, S-I~Itoh, K~Itoh and T.S.~Hahm},
\textit{ Plasma Phys. Control. Fusion} \textbf{47}, R35 (2005). \url{dx.doi.org/10.1088/0741-3335/47/5/R01}
\bibitem{Chang:95}
{Z.~Chang, J.D.~Callen, E.D.~Fredrickson, R.V.~Budny, C.C.~Hegna, K.M.~McGuire, M.C.~Zarnstorff, and TFTR group},
\textit{ Phys. Rev. Lett.} \textbf{74}, 4663 (1995). \url{dx.doi.org/http://dx.doi.org/10.1103/PhysRevLett.74.4663}
\bibitem{Carrera:86}
{R.~Carrera, R.D.~Hazeltine and M.~Kotschenreuther},
\textit{ Phys. Fluids} \textbf{29}, 899 (1986). \url{dx.doi.org/10.1063/1.865682}
\bibitem{McDevitt:06}
{C.J.~McDevitt and P.H.~Diamond},
\textit{ Phys. Plasmas} \textbf{13}, 032302 (2006). \url{dx.doi.org/10.1063/1.2177585}
\bibitem{Sen:09}
{A.~Sen, R.~Singh, D.~Chandra, P.~Kaw and D.~Raju},
\textit{ Nucl. Fusion} \textbf{49}, 115012 (2009). \url{dx.doi.org/10.1088/0029-5515/49/11/115012}



\bibitem{Ishizawa:07a}
{A.~Ishizawa and N.~Nakajima},
\textit{ Phys. Plasmas} \textbf{14}, 040702 (2007). \url{dx.doi.org/10.1063/1.2716669}
\bibitem{Ishizawa:07b}
{A.~Ishizawa and N.~Nakajima},
\textit{ Nucl. Fusion} \textbf{47}, 1540 (2007). \url{dx.doi.org/10.1088/0029-5515/47/11/016}
\bibitem{Militello:08}
{F.~Militello, F.L.~Waelbroeck, R.~Fitzpatrick, and W.~Horton},
\textit{ Phys. Plasmas} \textbf{15}, 050701 (2008). \url{dx.doi.org/10.1063/1.2917915}
\bibitem{Waelbroeck:09a}
{F.L.~Waelbroeck, F.~Militello, R.~Fitzpatrick, and W.~Horton},
\textit{ Plasma Phys. Cont. Fusion} \textbf{51}, 015015 (2009). \url{dx.doi.org/10.1088/0741-3335/51/1/015015}
\bibitem{Waelbroeck:09b}
{F.L.~Waelbroeck},
\textit{ Nucl. Fusion} \textbf{49}, 104025 (2009). \url{dx.doi.org/10.1088/0029-5515/49/10/104025}
\bibitem{Muraglia:09}
{M.~Muraglia, O.~Agullo, S.~Benkadda, X.~Garbet, P.~Beyer and A.~Sen},
\textit{ Phys. Rev. Lett} \textbf{103}, 145001 (2009). \url{dx.doi.org/10.1103/PhysRevLett.103.145001}
\bibitem{Muraglia:11}
{M.~Muraglia, O.~Agullo, S.~Benkadda, M.~Yagi, X.~Garbet and A.~Sen},
\textit{ Phys. Rev. Lett} \textbf{107}, 095003 (2011). \url{dx.doi.org/http://dx.doi.org/10.1103/PhysRevLett.107.095003}
\bibitem{Ishizawa:10a}
{A.~Ishizawa and N.~Nakajima},
\textit{ Phys. Plasmas} \textbf{17}, 072308 (2010). \url{dx.doi.org/10.1063/1.3463435}
\bibitem{Ishizawa:10b}
{A.~Ishizawa and P.H.~Diamond},
\textit{ Phys. Plasmas} \textbf{17}, 074503 (2010). \url{dx.doi.org/10.1063/1.3460346}
\bibitem{Ishizawa:13}
{A.~Ishizawa, S.~Maeyama, T.H.~Watanabe, H.~Sugama and N.~Nakajima},
\textit{ Nucl. Fusion} \textbf{53} 053007 (2013). \url{dx.doi.org/10.1088/0029-5515/53/5/053007}
\bibitem{Agullo:14}
{O.~Agullo, M.~Muraglia, A.~Poyé, S.~Benkadda, M.~Yagi, X.~Garbet, and A.~Sen},
\textit{ Phys. Plasmas} \textbf{21}, 092303 (2014). \url{dx.doi.org/10.1063/1.4894699}
\bibitem{Hu:14}
{Z.Q.~Hu, Z.X.~Wang, L.~Wei, J.Q.~Li and Y.~Kishimoto},
\textit{ Nucl. Fusion} \textbf{54}, 123018  (2014). \url{dx.doi.org/10.1088/0029-5515/54/12/123018}
\bibitem{Hill:15}
{P.~Hill, F.~Hariri, and M.~Ottaviani},
\textit{ Phys. Plasmas} \textbf{22}, 042308 (2015). \url{dx.doi.org10.1063/1.4919031}



\bibitem{Wilson:09}
{H.R.~Wilson and J.W.~Connor},
\textit{ Plasma Phys. Control. Fusion} \textbf{51}, 115007 (2009). \url{dx.doi.org/10.1088/0741-3335/51/11/115007}
\bibitem{Peeters:09}
{A.G.~Peeters, Y.~Camenen, F.J.~Casson, W.A.~Hornsby,A.P.~Snodin, D.~Strintzi and G.~Szepes},
\textit{ Comput. Phys. Comm.} \textbf{180}, 2650 (2009). \url{dx.doi.org/10.1016/j.cpc.2009.07.001}
\bibitem{Poli:09}
{E.~Poli, A.~Bottino and A.G.~Peeters},
\textit{ Nucl. Fusion} \textbf{49}, 075010 (2009). \url{dx.doi.org/10.1088/0029-5515/49/7/075010}
\bibitem{Poli:10}
{E.~Poli, A.~Bottino, W.A.~Hornsby, A.G.~Peeters, T.~Ribeiro, B.D.~Scott and M.~Siccinio},
\textit{ Plasma Phys. Control. Fusion} \textbf{52}, 124021 (2010). \url{dx.doi.org/10.1088/0741-3335/52/12/124021}
\bibitem{Hornsby:11}
{W.A.~Hornsby, M.~Siccinio, A.G.~Peeters, E.~Poli, A.P.~Snodin, F.J.~Casson, Y.~Camenen and G.~Szepesi},
\textit{ Plasma Phys. Control. Fusion} \textbf{53}, 054008 (2011). \url{dx.doi.org/10.1088/0741-3335/53/5/054008}
\bibitem{Siccinio:11}
{M.~Siccinio, E.~Poli, F.J.~Casson, W.A.~Hornsby and A.G.~Peeters},
\textit{ Phys. Plasmas} \textbf{18}, 122506 (2011). \url{dx.doi.org/10.1063/1.3671964}
\bibitem{Waltz:12}
{R.E.~Waltz and F.L.Waelbroeck},
\textit{ Phys. Plasmas} \textbf{19}, 032508 (2012). \url{dx.doi.org/10.1063/1.3692222}
\bibitem{Liu:14}
{Y.-H.~Liu, W.~Daughton, H.~Karimabadi, H.~Li, and S.P.~Gary},
\textit{ Phys. Plasmas} \textbf{21}, 022113 (2014). \url{dx.doi.org/10.1063/1.4865579}
\bibitem{Zarzoso:15}
{D.~Zarzoso, F.J.~Casson, W.A.~Hornsby, E.~Poli, and A.G.~Peeters},
\textit{ Phys. Plasmas} \textbf{22}, 022127 (2015). \url{dx.doi.org/10.1063/1.4908549}
\bibitem{Hornsby:15a}
{W.A.~Hornsby, P.~Migliano, R.~Buchholz, L.~Kroenert, A.~Weikl, A.G.~Peeters, D.~Zarzoso, E.~Poli and F.J.~Casson},
\textit{ Phys. Plasmas} \textbf{22}, 022118 (2015). \url{dx.doi.org/10.1063/1.4907900}
\bibitem{Hornsby:15b}
{W.A.~Hornsby, P.~Migliano, R.~Buchholz, D.~Zarzoso, F.J.~Casson, E.~Poli and A.G.~Peeters},
\textit{ Plasma Phys. Control. Fusion} \textbf{57}, 054018 (2015). \url{dx.doi.org/10.1088/0741-3335/57/5/054018}
\bibitem{Rutherford:73}
{P.H.~Rutherford},
\textit{ Phys. Fluids} \textbf{16}, 1903 (1973). \url{dx.doi.org/10.1063/1.1694232}



\bibitem{Dudson_2009_CPC_180}
{B.~Dudson, M.V.~Umansky, X.Q.~Xu, P.B.~Snyder and H.R.~Wilson},
\textit{ Comput. Phys. Comm.} \textbf{180}, 1467 (2009). \url{dx.doi.org/10.1016/j.cpc.2009.03.008}



\bibitem{Hazeltine_1985_PoF_28}
{R.D.~Hazeltine, M.~Kotschenreuther and P.J.~Morrison},
\textit{ Phys. Fluids} \textbf{28}, 2466 (1985). \url{dx.doi.org/10.1063/1.865255}; \textit{Phys. Fluids} \textbf{29}, 341 (1986).

\bibitem{Ishizawa:12}
{A.~Ishizawa, F.L.~Waelbroeck, R.~Fitzpatrick, W.~Horton and N.~Nakajima},
\textit{ Phys. Plasmas} \textbf{19}, 072312 (2012). \url{dx.doi.org/10.1063/1.4739291}
\bibitem{Fitzpatrick:05a}
{R.~Fitzpatrick and F.L.~Waelbroeck},
\textit{ Phys. Plasmas} \textbf{12}, 022308 (2005). \url{dx.doi.org/10.1063/1.1833391}
\bibitem{Fitzpatrick:05b}
{R.~Fitzpatrick, P.~Watson, and F.L.~Waelbroeck},
\textit{ Phys. Plasmas} \textbf{12}, 082510 (2005). \url{dx.doi.org/10.1063/1.2001644}
\bibitem{Fitzpatrick:05c}
{R.~Fitzpatrick and F.L.~Waelbroeck},
\textit{ Phys. Plasmas} \textbf{12}, 122511 (2005). \url{dx.doi.org/10.1063/1.2146983}

\bibitem{Hsu_1986_PoF_29}
{C.T.~Hsu, R.D.~Hazeltine and P.J.~Morrison},
\textit{ Phys. Fluids} \textbf{29}, 1480 (1986). \url{dx.doi.org/10.1063/1.865665}
\bibitem{Brizard:92}
{A.~Brizard},
\textit{ Phys. Fluids B} \textbf{4}, 1213 (1992). \url{dx.doi.org/10.1063/1.860129}
\bibitem{Scott_2000_PoP_7}
{B.~Scott},
\textit{ Phys. Plasmas} \textbf{7}, 1845 (2000). \url{dx.doi.org/10.1063/1.874007}



\bibitem{Hazeltine:87}
{R.D.~Hazeltine, C.T.~Hsu, and P.J.~Morrison},
\textit{ Phys. Fluids} \textbf{30}, 3204 (1987). \url{dx.doi.org/10.1063/1.866527}
\bibitem{Miyato_2004_PoP_11}
{N.~Miyato and Y.~Kishimoto},
\textit{ Phys. Plasmas} \textbf{11}, 5557 (2004). \url{dx.doi.org/10.1063/1.1811088}
\bibitem{Ishizawa_2007_PoP_14}
{A.~Ishizawa and N.~Nakajima},
\textit{ Phys. Plasmas} \textbf{14}, 040702 (2007). \url{dx.doi.org/10.1063/1.2716669}
\bibitem{Ishizawa_2013_PoP_20}
{A.~Ishizawa and F.L.~Waelbroeck},
\textit{ Phys. Plasmas} \textbf{20}, 122301 (2013). \url{dx.doi.org/10.1063/1.4838176}

\bibitem{Izacard_2011_PoP_18}
{O.~Izacard, C.~Chandre, E.~Tassi, and G.~Ciraolo},
\textit{ Phys. Plasmas} \textbf{18}, 062105 (2011). \url{dx.doi.org/10.1063/1.3591364}





\bibitem{Meneghini_2013_PFR_8}
{O.~Meneghini and L.~Lao},
\textit{ Plasma Fusion Res.} \textbf{8}, 2403009 (2013). \url{jspf.or.jp/PFR/PFR_articles/pfr2013S1/pfr2013_08-2403009.html}
\bibitem{Meneghini_2015_NF_IAEA}
{O.~Meneghini, S.P.~Smith, L.L.~Lao, O.~Izacard, Q.~Ren, J.M.~Park, J.~Candy, Z.~Wang, C.J.~Luna, V.A.~Izzo, B.A.~Grierson, P.B.~Snyder, C.~Holland, J.~Penna, G.~Lu, P.~Raum, A.~McCubbin, D.M.~Orlov, E.A.~Belli, N.M.~Ferraro, R.~Prater, T.H.~Osborne, A.D.~Turnbull and G.M.~Staebler},
\textit{ Nucl. Fusion} \textbf{55}, 083008 (2015). \url{dx.doi.org/10.1088/0029-5515/55/8/083008}






%

\end{thebibliography}
\end{document}